\documentstyle[epsfig]{article}
\newcommand{\beq}{\begin{equation}}
\newcommand{\eeq}{\end{equation}}
\newcommand{\be}{\begin{eqnarray}}
\newcommand{\ee}{\end{eqnarray}}

\def\fun#1#2{\lower3.6pt\vbox{\baselineskip0pt\lineskip.9pt
\ialign{$\mathsurround=0pt#1\hfil##\hfil$\crcr#2\crcr\sim\crcr}}}

\input{epsf}

\begin{document}

\title{The lightest scalar glueball}

\date{}
\author{ V.V. Anisovich}
\maketitle
\vskip 0.5cm
\begin{center}
{\it TO THE MEMORY OF YURI DMITRIEVICH PROKOSHKIN}
\end{center}
\vskip 0.5cm
\begin{abstract}
Recently performed investigations of meson spectra allow us to
determine the resonance structure for the waves  $IJ^{PC}=00^{++}$,
$10^{++}$, $02^{++}$, $12^{++}$ = $IJ^P=\frac12 0^+$
in the mass region up to 1900 MeV, thus establishing the  meson
multiplets $1^3P_0q\bar q$ and $2^3P_0q\bar q$.
Experimental data demonstrate that there are five
scalar/isoscalar states in this mass region. Four of them are $q\bar
q$ states, that is,  members of  the $1^3P_0q\bar q$ and $2^3P_0q\bar q$
nonets, while the fifth state is an extra one not accomodated by
$q \bar q$ systematics; it has the properties of the
lightest scalar  glueball. Analysis of the $00^{++}$-wave
performed within
the framework of the dispersion
relation technique allows us to
reconstruct the mixing of a pure gluonium  with neighbouring scalar $q
\bar q$ states belonging to $1^3P_0q\bar q$ and $2^3P_0q\bar q$
nonets: three scalar mesons share the gluonium state between each
other -- those are two comparatively narrow resonances  $f_0(1300)$
and $f_0(1500)$ and a broad resonance  $f_0(1530^{+90}_{-250})$. The
broad state is a descendant of the gluonium, keeping about 40-50\% of its
component.

\end{abstract}

\section{Introduction: retrospective view and the current state of
the problem            }

A great variety of the currently observed mesons and baryons represent
systems built of quarks. These are baryons, which are three--quark
systems $(qqq)$, and mesons, which are quark--antiquark bound states
 $(q\bar q)$. More than 20 years ago the problem  arose
\cite{fritz} whether additional hadrons exist which are built out of another
fundamental QCD particle, the gluon. An
intensive search for the glueball -- a
particle consisting of gluons -- was carried out throughout
these decades.

First evaluation of the glueball masses for different $J^{PC}$ was
done in the bag model \cite{1}. According to it, the lightest glueballs
are scalars and tensors, $0^{++}$ and $2^{++}$; then follow
pseudoscalar and  pseudotensor glueballs, $0^{-+}$ and $2^{-+}$.

Recently  considerable progress has been achieved in
Lattice QCD calculations. The UKQCD collaboration \cite{ukqcd} obtained the
following mass values for the lightest gluodynamical glueballs (i.e.
the glueball without quark degrees of freedom taken into account):
$$
m_G(0^{++})=1549\pm 53\, \mbox{MeV}, \qquad m_G(2^{++})=2310\pm 110\, \mbox{MeV},
\qquad m_G(0^{-+})=2332\pm 264\, \mbox{MeV}\,.  \eqno{(1.1)} $$
Systematic
errors are not included into the  values given in Eq. (1.1); they are
of the order of 100 MeV.

The IBM group obtained a slightly different value for the mass of the
lightest scalar glueball \cite{ibm}:
$$ m_G(0^{++})=1740\pm 71\,
\mbox{MeV}\,, \eqno{(1.2)} $$
The result of Ref. \cite{diego} is as follows:
$$
m_G(0^{++})=1630\pm 60 \pm 80 \, \mbox{MeV}\,, \qquad
m_G(2^{++})=2400\pm 10\pm 120 \, \mbox{MeV}\, .
\eqno{(1.3)}
$$
However, in Lattice calculations cited above the quark degrees of
freedom have not been taken into account, since the existing computing
facilities did not allow it. Quark degrees of freedom may
shift the position of the glueball mass noticeably. The dispersion
relation analysis
of meson spectra \cite{5,aas},
based on a restoration of the propagator matrix for
 scalar/isoscalar resonances,
shows that the mixing with $q\bar q$-states results in a mass shift
of the order of 100--300 MeV. It should be stressed that,
according to the
$1/N$ expansion rules \cite{thooft} ($N=N_f=N_c$, where $N_f$ and
$N_c$ stand for the light flavour and  colour numbers), the mixing
of a glueball with   $q\bar q$ states is not suppressed.

Experimental searches for the glueballs were particularly intensive during
the last decade. There exist  reactions where one could
expect an enhanced production of glueballs. Central hadron production
at the high--energy hadron--hadron collisions provides us
with an example of
such a reaction, because the particles in the central region are
produced in the transition {\bf pomerons} $\rightarrow$ {\bf hadrons}.
The Pomeron is a gluon--rich system, so one could expect to find among
secondary hadrons dominant production of
glueballs, while the
production of $q\bar q$-states is expected to be small.
However,  the data on
central hadron production, with statistics sufficient  to
perform a reliable partial wave analysis, are only now appearing. More
accessible for experimental study  appeared
to be another reaction,  which is
governed by the transition {\bf gluons} $\rightarrow$ {\bf hadrons};this
is, radiative decays $J/\psi\rightarrow\gamma+${\bf hadrons }. In
these decays, hadrons are formed by gluons created in   $c\bar c$
annihilation; therefore one may expect   dominant production of
glueballs
in this reaction. Experimental study of hadron spectra in
radiative $J/\psi$ decays has been carried out during 2 decades, and  is
still going on.  Experimental information accumulated at the beginning
of the 90's seemed to be rather discouraging, for in radiative
$J/\psi$ decays $q\bar q$ states have been strongly produced:
meson production branching ratios presented in the PDG
compilation \cite{pdg} show
a number of resonances produced
with similar probabilities, such as $\eta$, $\eta'$, $f_2(1270)$,
$f_2(1525)$, etc., which certainly are   $q\bar q$--dominant
 systems. Such a situation presents a dilemma:

{\bf (1)} The glueball does not exist; it is "an unfulfilled promise
of QCD" \cite{9}.

{\bf (2)} Glueball states are mixed strongly with the  $q\bar q$
mesons, so in experiments one observes just these mixed states.

Analysis of the $00^{++}$ wave \cite{5,aas} definitely supports
the second scenario.

Experimental data on the transition form factors
$\gamma\gamma^*(Q^2) \rightarrow \pi^0, \eta, \eta'$ \cite{10} provide
the following restrictions for probabilities to find the glueball
components in $\eta$ and  $\eta'$ mesons:
 $W_\eta\leq 8\%$, $W_{\eta'}\leq 20\%$ \cite{11}.
This means that in the  $q\bar q$ mesons observed in radiative
$J/\psi$ decay one could find a glueball component at the
level of 5-10\%. Hence,  the admixture of the $q\bar q$
component in the glueball should be considerably more, since the
glueball can mix with several  $q\bar q$ mesons. This qualitative
estimation agrees with that obtained in the framework of the $1/N$
expansion:  according to this, the glueball component in each
$q\bar q$ meson is of the order of $1/N_c$,
while  the  $q\bar q$ component
in the glueball is of the order of $N_f/N_c$ \cite{an}. Of course, it
should be stressed that some specific cases may differ from this
general evaluation, because the mixing  depends strongly on
the relative spacing of mixed levels.

If the scenario {\bf (2)} with strong mixing of the glueball and
 $q\bar q$ states is realized in Nature,
 the search for the glueball is laborious and difficult work involving
the identification of mesons and their systematics. Naive
expectations, such as a study of the gluon--rich reactions with the
 purpose of seeing direct glueball production or, conversely, a study
of hadron production in $\gamma \gamma \to q\bar q$-induced
reactions with the hope of excluding gluon-rich hadrons,
 cannot be expected to succeed.

The main channel of radiative  $J/\psi$ decays, as deduced from
experimental data, is the production of  broad hadron clusters.
The production of these clusters may be viewed as a direct signal of
strong mixing between the glueball  and $q\bar q$ mesons.
What happens is that, through mixing, one resonance accumulates the widths
of other resonances. This
effect has been first observed
 in \cite{12}, where the low--energy part of
the spectrum of the $00^{++}$ wave has been analysed; this
effect has been investigated in detail  \cite{5,aas}. When the
two resonances mix completely with each other, one of them
 gets almost the
whole width  $\Gamma_1+\Gamma_2$, while the width of the other one tends
to zero. In the case of an "ideal" mixing of three resonances, the width
of one of them accumulates the widths of the other two,
$\Gamma_1+\Gamma_2+\Gamma_3$, and the widths of the others tend to
zero. In reality, when the scalar glueball mixes with  neighbouring
states, there occurs a qualitatively similar  effect; that is, a glueball
situated among the scalar $q\bar q$ states mixes with them and
accumulates a considerable part of their widths. From this point of
view, the appearance of a broad resonance which is the glueball
descendant is an inevitable consequence of the mixing. The broad
resonance must be a neighbour of comparatively narrow resonances, which
are the descendant of pure $q\bar q$ states; the broad resonance
contains a considerable glueball admixture. The analysis of the
$00^{++}$ wave in the mass range 1200--1800 MeV, based on the dispersion
relation representation,  reconstructs just this
 picture of the lightest scalar glueball mixing with $q\bar q$
 members of multiplets $1^3P_0$ and $2^3P_0$.  One may
predict that such a scenario of mixing is common for all  low--lying
 glueballs.

Thus, the strong mixing of $q\bar q$ states with a gluonium
does not allow easy identification of the glueball. In this case the only
reasonable strategy is to study the systematics of all resonances in terms of
$q\bar q$ multiplets. The extra states which do not fit into
$q\bar q$ systematics should be regarded as candidates for the
glueballs or other exotic mesons. This investigation program has been
declared in \cite{an},  and at the same time the first steps have been
made in carrying it out: in Ref. \cite{12} the  $K$-matrix
analysis has been performed for the low--energy part of the wave
$IJ^{PC}=00^{++}$.

Detailed analysis of meson states in the region  1000-2000 MeV was
possible due to the huge sample of experimental data collected in the
latest decade by Crystal Barrel and GAMS Collaborations.
The Crystal Barrel Collaboration has  high--precision data on the
production of three neutral mesons in the reaction
 $p\bar p$ annihilation at rest,
$$ p\bar p\;(at\; rest)\rightarrow\pi^0\pi^0\pi^0\,,\qquad
\pi^0\pi^0\eta\,,\qquad\pi^0\eta\eta\,,
\eqno{(1.4)}
$$
with the event numbers 720000 for $(\pi^0\pi^0\pi^0)$,
280000 for $(\pi^0\pi^0\eta)$ and 185000 for $(\pi^0\eta\eta)$.
The data on the reaction  $p\bar p\; (at \;rest) \to \pi^0\pi^0\pi^0$,
with somewhat lower statistics were published in  1991 \cite{13}.
However, the first fits of the spectra did not provide a correct
identification of scalar resonances, for certain special
features of the three--particle decay were not
taken into account. A critical analysis of the situation has
been made in \cite{14,PR94}, where it was shown that a resonance near
 1500 MeV, which had earlier been identified as  a tensor one,
$AX_2(1520)$, actually is a scalar resonance.
Re-analysis of the reactions (1.4) within the $T$-matrix formalism
 performed together with the Crystal Barrel Collaboration has fixed the
existence of new scalar resonances:  $f_0(1500)$ \cite{15} and
$a_0(1450)$ \cite{16}.  In addition, in \cite{14,PR94,15} a noticeable
production of the resonance $f_0(1360)$, with a half-width equal to 130
MeV, weas identified,
although at that time it was not quite clear whether this was a newly
 observed resonance  or a fragment of the broad resonance
 $\epsilon (1300)$, which  was widely discussed during the latest decades.
Later on, after having performed the $K$-matrix analysis for larger
samples of data, it became clear that in this mass region
there are two resonances -- a comparatively narrow one, $f_0(1360)$, and a
 rather broad one , $f_0(1530^{+90}_{-250})$.

In the first stage of the investigation, the fitting to data was
done in the framework of the $T$-matrix technique. The reason was
obvious: the $T$-matrix representation of the amplitude is simpler for
fitting to data; the advantages of the $K$-matrix approach reveal
themselves only when there exists information on all possible
channels of the reaction. In the mass region  1000--1500 MeV, there are
the following channels in the $00^{++}$ wave:  $\pi\pi$, $K\bar K$,
$\eta\eta$  and  $\pi\pi\pi\pi$, while in the region above 1500 MeV
the channel $\eta\eta'$ becomes important. It was obvious that the
application of the sophisticated $K$-matrix technique for fitting to a
limited number of channels (1.4) would lead to some ambiguities.

The discovery of the resonance  $f_0(1500)$ immediately gave rise
to the hypotheses of its close relation to the lightest scalar
glueball, and the possibility of such a relation was
stressed in  \cite{PR94,15}.  In the following papers
\cite{an,ac,clofer,wein,genov}, several schemes have been suggested for
the mixing of the lightest scalar glueball with the neighbouring
 $q\bar q$-states. However,  all these schemes did not take
account of  special features of the mixing which are due to the transition
of a resonance into real mesons, although just these transitions, as was
shown in a specified $K$-matrix analysis, determine the structure
of the $00^{++}$-wave around 1500 MeV.

At the next stage of the $00^{++}$ wave analysis, the GAMS
data on the spectra  $\pi^0\pi^0$, $\eta\eta$ and $\eta\eta'$ have been
included; these were obtained in the reactions \cite{G1,G2,G3}:
$$
\pi^-p\rightarrow n\pi^0\pi^0\,,\qquad n\eta\eta\,,\qquad n\eta\eta'\,,
\eqno{(1.5)} $$
together with the data of the  CERN-M\"{u}nich collaboration
\cite{cern}:
$$ \pi^-p\rightarrow n\pi^+\pi^- \eqno{(1.6)} $$
and BNL \cite{bnl} group:
$$ \pi\pi\rightarrow K\bar K\,.  \eqno{(1.7)}
$$
Simultaneous analysis of the whole  data sample (1.4)-(1.7) was
carried out in Refs. \cite{12, km1500,km1900}, in the framework of
the $K$-matrix technique; in this way the range of masses under
 investigation and the number of channels covered by the $K$-matrix fit
of the $00^{++}$ amplitude gradually increased.

The first investigation \cite{12} was done
in the region of invariant meson energies $\sqrt{s}\leq 1100$ MeV
for the two channels only, namely, $\pi\pi$ and $K\bar K$,
 In this analysis an  observation, which
became henceforth important, was made: the transitions which are
responsible for the decay of meson states are also responsible for a
strong mixing of these states.  Moreover, the masses of mixed states
differ essentially from the primary ones. These "primary mesons" were
called in \cite{12} "bare mesons", in contrast to physical states,
 for which the cloud of real particles, $\pi\pi$
and $K\bar K$, plays an important role in their formation.  The masses
of bare states are defined as the $K$-matrix poles.
The above--mentioned  accumulation of widths of the primary states by one
of the resonances due to mixing, was also observed in Ref.
\cite{12}.

As the next step, the $K$-matrix analysis was extended  to
1550 MeV \cite{km1500}, with additional channels
$\eta\eta$ and $\pi\pi\pi\pi$ included.
The channel $\pi\pi\pi\pi$ is
rather important for the correct description of spectra from
1300 to 1600 MeV, since $\sigma(\pi\pi\to\pi\pi\pi\pi)/
\sigma(\pi\pi\to\pi\pi)$ is of the order of  0.5 at 1300 MeV and about
1.5 at 1500 MeV \cite{19}.  The use of channels
  $\pi\pi$, $K\bar K$ and $\eta\eta$ provides an opportunity to
perform the $q\bar q$ classification of bare  $00^{++}$ states,
$f_0^{bare}$, below 1600 MeV \cite{km1500}.
The point is that  $q\bar q$-meson decays   go to the  new $q\bar q$-pair
via the production of intermediate gluons. According to the rules
of the $1/N$ expansion, the main contribution to the decay coupling
constant comes from planar diagrams. When an isoscalar  $q\bar q$-
meson disintegrates into two pseudoscalar mesons $P_1P_2$, namely,
$$
\pi\pi\,,\qquad K\bar K\,,\qquad\eta\eta\,,\qquad\eta\eta'\,,
\qquad\eta'\eta'\,, \eqno{(1.8)}
$$
the coupling constants can be determined,  up to a
common factor, by  two factors. The first is the quark content of the
$q\bar q$-meson
$$ q\bar q=n\bar n\cos\phi+s\bar s\sin\phi \,, \eqno{(1.9)}
$$
where $n\bar n=(u\bar u+d\bar d)/\sqrt{2}$. The second is the
parameter $\lambda$, which characterises the relative probability to
produce nonstrange and strange quarks by gluons in  soft processes:
$$
u\bar u:d\bar d:s\bar s=1:1:\lambda\,.  \eqno{(1.10)} $$
Experimental data provide the following values for this parameter:
$\lambda\simeq 0.5$ \cite{ahkm} in central hadron production in
hadron--hadron high--energy collisions,
 $\lambda=0.8\pm 0.2$ \cite{klempt} for the decay of tensor mesons and
 $\lambda=0.6\pm 0.1$ \cite{a95,az} for  the ratio of yields of
 $\eta$ and $\eta'$ mesons in the decays
$J/\psi\rightarrow\gamma\eta/\gamma\eta'$.

Coupling constants for the decay
$q\bar q\rightarrow P_1P_2$ into channels (1.8), which are defined by
the leading planar diagrams in the $1/N$ expansion, may be presented as
$$
g(q\bar q\rightarrow P_1P_2)=C_{P_1P_2}(\phi,\lambda)g^L\,,
\eqno{(1.11)}
$$
where  $C_{P_1P_2}(\phi,\lambda)$ is a wholly calculable
coefficient depending on the mixing angle  $\phi$ and parameter
$\lambda$; $g^L$ is a common factor describing the unknown
dynamics of the process. Therefore, experimental investigation of
resonance decays into  channels (1.8) allows us to reconstruct the quark
content of the state (i.e. its mixing angle  $\phi$), thus making it
possible to establish the meson systematics.

However, on the basis of the decay constant analysis, it is impossible
to determine unambigously whether  we deal with a $q\bar q$-meson
or with the glueball. The reason is that the glueball decay is
a two--stage process, with a subsequent production of  two
$q\bar q$ pairs. After the production of the first  $q\bar q$
pair, in the intermediate state there exists a
 $q \bar q$ system with the  following $n\bar n/s\bar s$ content:
$$ n\bar
n\cos{\phi_{Glueball}}+s\bar s\sin{\phi_{Glueball}}  \; ,  \; \; \; \;
\; \tan{\phi_{Glueball}}=\sqrt{\frac{\lambda}{2}}\,.
\eqno{(1.12)}
$$
For $\lambda=0.45-0.80 $ the mixing angle is
$\phi_{Glueball} $ =$25^\circ-32^\circ $. At the second stage,
the intermediate $q\bar q$ state
(1.12) turns into the $P_1 P_2 $ mesons; this
means that the relations between glueball coupling constants are the
same as for the decay of the  $q\bar q$ meson with
$\phi $ = $\phi_{Glueball} $.

Analysis of the $\pi\pi$, $K\bar K$ and $\eta\eta$ spectra performed in
\cite{km1500} proved that in the region below 1600 MeV there are four
scalar/isoscalar states, and only one of them is an
 $s\bar s$-dominant state. Since each of the $^3P_0q\bar q$
multiplets contains  two $I=0$ states, which refer to two flavour
combinations $n\bar n$ and $s\bar s$, then, as a result of the
analysis of \cite{km1500}, the following dilemma becomes apparent:

{\bf 1.} In the region  1000-1800 MeV there are three  $^3P_0q\bar q$
nonets:  the basic one, $1^3P_0q\bar q$, and two  radial excitations,
$2^3P_0q\bar q$ and $3^3P_0q\bar q$. In this case, there should exist
two $s\bar s$-dominant scalar mesons in the region 1600-1800 MeV.

{\bf 2.} At 1600-1800 MeV, there is only one $s\bar s$-dominant
state. Then, one of the three mesons from the region 1200-1600 MeV is an
extra one from the point of view of $q\bar q$-systematics, and it
should be considered as a candidate for  an exotic meson: the ratios of
couplings to the channels  (1.8) found in \cite{km1500} provide the
basis to consider  it as the lightest scalar glueball.

Thus, after carrying out the analysis
of Ref. \cite{km1500}, the immediate task
was to extend the $K$-matrix analysis of the $00^{++}$ wave to the
region 1600--1900 MeV. Such an extension suggested the inclusion of
the  $\eta\eta'$ channel into the fitting procedure; this has been done
in \cite{km1900}, where the  $K$-matrix analysis has been
performed in the mass region 500-1900 MeV, with the next
five channels taken into consideration:
  $\pi\pi$, $K\bar K$, $\eta\eta$, $\pi\pi\pi\pi$, $\eta\eta'$.
It was shown that in the range 1600-1900 MeV there exists only
one $f_0$-meson with a  dominant $s\bar s$
component, hence the analysis \cite{km1900} confirmed the case
{\bf 2}. In this way, it was also shown that there are two
variants for fixing  the scalar glueball:

{\bf Solution I.} Two bare states, $f_0^{\rm bare}(720\pm 100)$ and
$f_0^{\rm bare}(1260\pm 30)$, are members of the multiplet $1^3P_0q\bar
q$, and $f_0^{\rm bare}(720)$ is the  $s\bar s$-rich state, with
$\phi(720)=-69^\circ\pm 12^\circ$.  The bare states $f_0^{\rm
bare}(1600\pm 50)$ and $f_0^{\rm bare}(1810\pm 30)$ are members of the
$2^3P_0q\bar q$ nonet , and $f_0^{\rm bare}(1600)$ is dominantly
$n\bar n$-state, with $\phi(1600)=-6^\circ\pm 15^\circ$. The state
$f_0^{\rm bare}(1235\pm 50)$ is superfluous from the point of view of
the  $q\bar q$ classification; its coupling constants satisfy the
 ratios relevant to gluonium. Therefore, this state may be considered
as a candidate for the lightest scalar glueball.

{\bf Solution II.} The basic scalar nonet is the same as in Solution I.
The members of the next nonet, $2^3P_0q\bar q$, are as follows:
$f_0^{\rm bare}(1235\pm 50)$  and $f_0^{\rm bare}(1810\pm 30)$. Both
these states contain a considerable admixture of the $s\bar s$
component:  $\phi(1235)=42^\circ\pm 10^\circ$  and
$\phi(1810)=-53^\circ\pm 10^\circ$.  The state $f_0^{\rm bare}(1560\pm
30)$ is an extra one from the point of view of $q\bar q$
systematics and it may be regarded as a good candidate for the
lightest scalar glueball.

The existence of  two variants corresponds to the impossibility to
answer unambiguously,  on the basis of the information
on the decay channels (1.8),  if we
deal with the glueball or  $q\bar q$-meson with the mixing angle $\phi
$ in the region $25^\circ-32^\circ $, as was stressed above.

Both $K$-matrix solutions, I and II, lead to practically  identical
positions of the amplitude poles in the complex mass plane.
The amplitude has five poles:
$$
\begin{array}{cl} Resonance & Pole\; position (in \;MeV) \\ ~& ~ \\
f_0(980) & 1015\pm15-i(43\pm8)\\ f_0(1300) &
1300\pm20-i(120\pm20)\\ f_0(1500) & 1499\pm8-i(65\pm10) \\ f_0(1750) &
1750\pm30-i(125\pm70)\\ f_0(1530^{+90}_{-250}) &
1530^{+90}_{-250}-i(560\pm140)\ .  \end{array} \eqno{(1.13)} $$
The broad resonance  $f_0(1530^{+90}_{-250})$ is not a new object in
meson physics: this is that  one which was
called $\epsilon (1300)$. A large width of  $f_0(1530^{+90}_{-250})$ is
due to the accumulation of widths of neighbouring resonances.

Reliable and unambigous identification of the scalar glueball must be
based upon the complete reconstruction of the multiplets  $1^3P_0q\bar
q$ and $2^3P_0q\bar q$. Each of these nonets consists of two
scalar/isoscalar states $f_0$, one isovector/scalar state $a_0$, and
the scalar kaon $K_0$. As was stated above, it is reasonable to perform
the nonet classification of highly--excited  $q\bar q$-states in terms
of bare states, which do not contain clouds of real  mesons. The
analysis \cite{km1900} fixed four $f_0^{bare}$ mesons,
which are necessary for the construction of two nonets;  the two lightest
isotriplet resonances, $a_0(980)$ and $a_0(1450)$, are also known, see
\cite{pdg}.  A full $K$-matrix analysis of the $10^{++}$ wave,
\cite{fullkm}, provided the following resonance masses:  $$ a_0(980)
\to (988\pm6)-i(46\pm10)\; \mbox{MeV}, \; a_0(1450) \to
(1565\pm30)-i(146\pm20)   \;  \mbox{MeV}  .  \eqno{(1.14)} $$
It should be pointed out that in the PDG compilation \cite{pdg}
the mass of the second resonance is too low by about 100 MeV.

Corresponding bare states are as follows:
$$
a_0^{\rm bare}(964\pm16), \; \; a_0^{\rm bare}(1670\pm70).
\eqno{(1.15)}
$$

Identification of scalar resonances as members of the
 $1^3P_0q\bar q$ and   $2^3P_0q\bar q$ always raised problems.
Namely, according to \cite{pdg,kpi}, the masses of the two lightest kaons
are $1429\pm 4\pm 5$ MeV  and $1945\pm 10\pm 20$ MeV;
these are noticeably higher than the average masses of other mesons which are
candidates for the scalar--nonet members. This high position on
the mass scale of the scalar kaon, $K_0(1430)$, gave impetus to models
where the basic $1^3P_0$ $q\bar q$-multiplet was fixed in the region
1350-1500 MeV, and the resonances $f_0(980)$ and $a_0(980)$
were considered as exotic states --- hadron molecules
\cite{21}, multiquark bags \cite{22}, or minions \cite{bonn,minion}.

In Ref. \cite{kmkpi} the $K$-matrix re-analysis of the $S$-wave
$K\pi$-spectrum was carried out to
determine the $K_0^{\rm bare}$. Another reason to re-analyse it was
as follows. In Ref. \cite{kpi} the $K\pi$-spectra have been
investigated in  two separate mass regions,  820-1580 MeV and
1780-2180 MeV, but the mass region  1580-1780 MeV was not included into the
analysis of the $K\pi$ amplitude.  Our  experience in fitting the
$00^{++}$-wave \cite{km1900} teaches us that separate consideration of
different mass regions leads to the loss of certain information. In
order to get a full picture, a simultaneous fit is needed;
moreover, at 1580-1780 MeV there is a rapid change of the amplitude.

As follows from the $K$-matrix fit of the  $(IJ^P =\frac12 0^+)$
wave \cite{kmkpi}, for a good description of the $K\pi$-spectrum in
the region 800-2000 MeV at least two $K_0$-states are necessary.
Correspondingly, the $\frac12 0^+$-amplitude of this minimal solution
has poles near the physical region on the 2nd sheet (under the
$K\pi$-cut) and on the  3rd sheet (under $K\pi$- and $K\eta'$-cuts), at
 the following complex masses:
 $$ (1415\pm 30)-i(165\pm 25)
\;\mbox{MeV} , \qquad \; (1820\pm 40)-i(125\pm 35) \;\mbox{MeV}.
\eqno{(1.16)} $$
The $K\eta'$ threshold, being in the vicinity of the resonance
(at 1458 MeV), strongly influences the $\frac12 0^+$
amplitude, so the lowest  $K_0$-state has a second pole which is
located  above the $K\eta'$-cut, at  $M=(1525\pm 125)-i(420\pm 80)$
 MeV:  the situation is analogous to that observed for the
 $f_0(980)$-meson, which also has a two--pole structure of the
amplitude due to the  $K\bar K$-threshold.  The $K\eta$ channel weakly
influences the $\frac{1}{2} 0^+ \; K\pi $ amplitude:   experimental
data \cite{kpi} prove it as well as the quark combinatoric rules.

The minimal solution contains two $K_0^{\rm bare}$ states:
 $$
K_0^{\rm bare}(1200^{+60}_{-110})\,,
\qquad K_0^{\rm bare}(1820^{+40}_{-75})\,.
\eqno{(1.17)}
$$
In this minimal solution, the lightest scalar bare kaon appears
200 MeV lower than the amplitude pole, and this latter
circumstance makes it easier to build the basic scalar nonet, with
masses in the range 900-1200 MeV.

It is worth noting that the  $K\pi$ spectra also allow solutions with
three poles, with much better $\chi ^2$; still, for these solutions
the lightest kaon state, $K_0^{\rm bare}$, does not leave the range
  900-1200 MeV. In the tree--pole solution
$$
K_0^{bare}(1090\pm 40)\, ,
\qquad K_0^{bare}(1375^{+ 125}_{-40})\, ,
\qquad K_0^{bare}(1950^{+70}_{-20})\, ,
\eqno{(1.18)}
$$
and the $K\pi$-amplitude has the poles:
\be
\mbox{II sheet}&\;M=998 \pm 15-i\;(80 \pm 15)&\;\mbox{MeV}
\nonumber \\
\mbox{II sheet}&\;M=1426 \pm 15-i\;(182 \pm 15)&\;\mbox{MeV}
\nonumber \\
\mbox{III sheet}&\;M=1468 \pm 30-i\;(309 \pm 15)&\;\mbox{MeV}
\nonumber \\
\mbox{III sheet}&\;M=1815 \pm 25-i\;(130 \pm 25)&\;\mbox{MeV}.
\nonumber
\ee
$$
\eqno{(1.19)}
$$
The state $ K_0^{bare}(1375^{+ 125}_{-40}) $, being near the
$K\eta'$ threshold, results in doubling the amplitude poles
around 1400 MeV. It should be underlined that masses of the
lightest bare kaon states obtained by the two-- and three--pole
solutions coincide within the errors.

The $K$-matrix fit of the $\frac{1}{2} 0^+$ wave makes it possible to
complete, in terms of bare states,  the construction of the two lowest
scalar nonets. In line with the result  for  the $00^{++}$ wave
\cite{km1900}, where  two solutions
for an extra state (candidate for the glueball) were found,
 there are   two variants for the nonet classification of scalar mesons.
The basic  $1^3P_0q\bar q-$ nonet is the same for both variants:
 $$ a_0^{\rm bare}(960\pm 30)\; , \; f_0^{\rm bare}(720\pm
 100)\; , \; f_0^{\rm bare}(1260^{+100}_{-30})\; , \; K_0^{\rm
 bare}(1200^{+90}_{-150}) \; .  \eqno{(1.20)} $$
It should be particularly stressed that the wave function
  $f_0^{\rm bare}(720)$  in the flavour space is close to the
octet one; indeed, $\phi (720)=-69^o\pm 12^o$, while
$\phi_{octet}=-54.7^o $.  Correspondingly,  $  f_0^{\rm bare}(1260)$
is close to the flavour singlet. A similar situation is observed
in the pseudoscalar sector where the flavour wave functions
$\eta$ and $\eta'$ are close to the octet and singlet ones.
It is even more analogous, if one takes into consideration that the
mass diffrence of isoscalar states in these sectors coincide with
each other,
and the scalar masses are not much larger than corresponding masses of
pseudoscalars, $m_{scalar}- m_{pseudoscalar} \simeq (200\pm 100)$ MeV.
Such coincidences clearly point towards parity degeneration of the
interaction forces in isoscalar channels.

Thus, one may conclude: the basic nonet of scalar mesons is uniquely
fixed by the  $K$-matrix fit of meson sspectra. It is located rather
low on the mass scale, in the range  750-1250 MeV.  Here, at  mass
values below 1200 MeV, there is no room for exotic states.

The $2^3P_0q\bar q$ nonet contains the following states in
Solution I:
 $$ a_0^{\rm bare}(1640\pm 40) \; , \;
f_0^{\rm bare}(1600\pm 50) \; , \; f_0^{\rm bare}(1810^{+30}_{-100}) \;
, \nonumber $$ $$ K_0^{\rm bare}(1375^{+ 125}_{-40}) \; or \;
K_0^{\rm bare}(1820^{+40}_{-60}) \; .
\eqno{(1.21)}
 $$
An extra state with respect to the  nonet classification is
$f_0^{\rm bare}(1235^{+150}_{-30})$.

In Solution II the $2^3P_0q\bar q$ nonet looks like:
$$
a_0^{\rm bare}(1640\pm 40) \; , \;
f_0^{\rm bare}(1235^{+150}_{-30} ) \; , \;
f_0^{\rm bare}(1810^{+30}_{-100}) \; , \;
\nonumber
$$
$$
K_0^{\rm bare}(1375^{+ 125}_{-40}) \;or \;
K_0^{\rm bare}(1820^{+40}_{-60}) \; .
\eqno{(1.22)}
 $$
In this solution an extra state is $f_0^{\rm bare}(1600\pm 50)$;
once again it should be stressed that the mass of this state appears
just in the mass region where Lattice calculations
for the mass of the lightest scalar gluonium point; also the
coupling constants with meson channels agree with the quark
combinatorics ratios for the gluonium decay.

Immediately after performing the $K$-matrix analysis in the
range up to 1900 MeV, the problem of presentation of the $00^{++}$
amplitude as a dispersion integral
 has been raised. The dispersion $N/D$
representation correctly restores  analytic properties of partial
amplitude in the whole complex  $s$-plane. In addition, and this the
principal point, within the dispersion representation it is possible to
reconstruct  the propagator matrix, thus evaluating the
mixing  of the $q\bar q$-states and the glueball, and then to restore
correctly the gluonium mass. The dispersive $N/D$-description of the
wave $00^{++}$ has been performed in Refs.  \cite{5,aas}: in Ref.
\cite{5} the region 1200-1700 MeV where three scalar/isoscalar states
are located, has been studied;  then, in Ref.  \cite{aas} the region
under investigation has been extended  to 1900 MeV, with the fourth
state, $f_0(1780)$, included into consideration.

The results of the  $N/D$-representation of the $00^{++}$ wave allowed
us to draw the picture of mixing for the lowest scalar gluonium: it
mixes with the two neighbouring $q\bar q$ states ---
members of the multiplets $1^3P_0$ and $2^3P_0$, and the resonance--
descendant of a pure glueball accumulates large parts of the widths of the
neighbouring resonances, being transformed into the broad state
 $f_0\left(1530{+90\atop -250}\right)$ .

It should be emphasized that the state  $f_0^{\rm bare}$,
which was found in the $K$-matrix fit,
does not explicitly describe the gluodynamic glueball, for the state
  $f_0^{\rm bare}$ contains nongluonic degrees of freedom
related to  real parts of the loop diagrams (imaginary parts are
responsible for the decay process). The dispersion relation $N/D$ method
is able to restore the real and imaginary parts of the loop diagrams,
thus providing a complete picture of  mixing, so it also restores
the mass of gluonium. In Solution I it is equal to
$$
  m_{gluonium}=1225\; \mbox{MeV}\; , \eqno{(1.23)} $$
and in Solution II:  $$
m_{gluonium}=1633\; \mbox{MeV}\; .  \eqno{(1.24)} $$
The mass
$m_{gluonium}=1633$ MeV (Solution II) agrees well with the mass values
of the lightest scalar glueball obtained in Lattice calculations.

It is rather striking that both solutions obtained in the dispersive
technique provide practically the same structure for the
$00^{++}$ wave and the quark--gluon content of physical
resonances:  in both solutions the broad resonance,
$f_0\left(1530{+90\atop -250}\right)$, is a descendant of the
gluonium, keeping about 40-50\% of its component, while the
remnant part of the gluonium is  shared between $f_0(1300)$
and $f_0(1500)$.  From this point of view, the structure of resonances
in the range 1300-1600 MeV is uniquely solved.

The formation of  the broad state which is seen in the $00^{++}$
wave raises a question about the presence of such effects in other
waves as well, for it is reasonable to believe that exotic mesons
(glueballs and hybrids) with other quantum numbers can also
afford a width accumulation. Because of that, the search for other
exotic mesons must be inseparably linked with the study of broad
resonances.

There appears another problem which is not less intriguing: the broad
resonance, after having absorbed the widths of its
neighbouring resonances, plays the role of a locking state, since it
prevents the decay of neighbouring states with the same quantum
numbers. This means that the broad resonance actually plays the role of
a dynamic barrier for the nearby states. How does this dynamic barrier
relate to the confinement barrier? --- Only detailed investigations
of  broad resonances in other waves could answer this question.

\section{$K$-matrix and the dispersive $N/D$-representation of the
scattering amplitude}

In this Section a brief review is done  for the technique used in the
analysis of meson spectra. Namely,  analytic properties of the
amplitude are discussed together with the connection of the dispersion
$N/D$ representation to the  $K$-matrix approach. The role of short
and large distances in the formation of meson spectra
under investigation is
also discussed, and in this connection a notion "bare state" is
introduced. The quark combinatoric relations between the decay
couplings of the glueball to meson channels are considered in comparison
with similar relations for $q\bar q$-states: these relations provide
the basis for the nonet systematics of mesons.

\subsection{Scattering amplitude, $T$-matrix and $K$-matrix}

Basing on a simple example, let us get over
the terminology and
notations used for the analysis of meson spectra.

In terms of the wave function, which expresses relative movement of
two spinless particles, the scattering at large distances is
described by the incoming plane and outgoing spherical waves, with
the coefficient $f(\theta)$  related to the
partial amplitudes as follows:
$$
f(\theta)=\frac{1}{2ik}\sum_{\ell=0}^{\infty}(2\ell+1)P_\ell
(\cos\theta)\left[e^{2i\delta_\ell(k)}-1\right]\,.
\eqno{(2.1)}
$$
This formula is written for the one-channel scattering in absence of
absorbtion ($k$ is the momentum of relative movement,
$\ell$ is the angular momentum and
$\theta$ is the scattering angle). The $T$-matrix element is determined
by the scattering phase shift $\delta_\ell$:
$$T_\ell=\frac{1}{2i}\left[e^{2i\delta_\ell}-1\right]=
e^{i\delta_\ell}\sin\delta_\ell\, .\eqno{(2.2)}$$
For the investigation of analytic properties, it is suitable to
use an amplitude with other normalization:
$$A_\ell=\frac{1}{2i\rho(k)}\left[e^{2i\delta_\ell}-1\right]\,,
\eqno{(2.3)}$$
where $\rho(k)$ is the invariant two-particle phase space factor:
 $$ \rho(k)=\int d\Phi(P;k_1,k_2)\,, \qquad
d\Phi(P;k_1,k_2)=\frac 12\frac{d^3k_1}{(2\pi)^32k_{10}}
\frac{d^3k_2}{(2\pi)^32k_{20}}(2\pi)^4\delta^{(4)}(P-k_1-k_2)\,.
\eqno{(2.4)}
$$
Invariant phase space factor
is determined by three four-momenta:  the total
momentum of scatterred particles,  $P$, with $P^2=s$, and the momenta of
particles 1 and 2, $k_1$ and $k_2$, respectively.
For equal masses of particles 1
and 2, we get:
$$\rho(k)=\frac{k}{8\pi\sqrt{s}}\,,\qquad k=\sqrt{\frac{s}{4}-m^2}\,.
\eqno{(2.5)}$$
The $K$-matrix representation of the amplitude $A_\ell$ reads:
$$A_\ell=\frac{K_\ell(k^2)}{1-i\rho(k)K_\ell(k^2)}.  \eqno{(2.6)}$$
$K_\ell$ is real in the physical region;
the imaginary part of the
amplitude is explicitly written in Eq. (2.6). In addition,  $K_\ell$
as a function of the invariant energy squared $s$ is analytic
near the threshold singularity, $s=4m^2$: a singular
term is  singled out,  it is explicitly given by the two-particle
phase space factor $\rho$.

In presence of absorbtion,
the scattering with an absorbtion is described by
 the absorbtion coefficient $\eta_\ell$ inserted to the partial wave
expansion (2.1):
$$\left[e^{2i\delta_\ell}-1\right]\rightarrow\left[\eta_\ell
e^{2i\delta_\ell}-1\right]\;. \eqno{(2.7)}$$
Here $0\leq\eta_\ell\leq 1$; the case  $\eta_\ell=0$ corresponds to
full absorbtion.

It is suitable to display the  energy--dependent  amplitude
$T_\ell$ on the Argand--diagram, which is an appropriate
instrument for searching resonances. The $T$-matrix element at
fixed $k$ (or $s$) corresponds to the
point on the plane $(Re \,T_\ell, Im \,T_\ell)$.
As a function of $k$, it draws a trajectory on the
circle with the radius equal to 1/2 and
the  centre in the point  $0,i/2$. In
the inelasticity case, the trajectory $T_\ell$ enters the internal
part of the circle.

The $K$-matrix representation of the amplitude with an absorbtion
requires fixing inelastic channels. Consider the
inelasticity  occurring
due to another two-particle channel; we denote these channels
 by indices  1 and 2.  Then the elastic scattering
amplitude $1+1'\to 1+1' $ (denoted  as $A_{11}$, index
$\ell$ is omitted) can be presented in the form of Eq.
(2.6):
$$A_{11}=\frac{K(k^2)}{1-i\rho_1K(k^2)}\,.\eqno{(2.8)}
$$
However, the block $K(k^2)$  has an imaginary part above the
threshold of the second channel:
$$K(k^2)=K_{11}+i\frac{K_{12}\rho_2K_{21}}{1-i\rho_2K_{22}}\,.
\eqno{(2.9)}$$
Here $\rho_2$ is phase space factor of the second channel $2+2' $,
and matrix elements $K_{11}$,   $K_{12}=K_{21}$ and $K_{22}$ are real
functions of $k^2$ in the
physical region.
Threshold singularities of the channels 1 and 2, which are located at
$s=(m_1+m'_1)^2$ (threshold of the channel 1) and at $s=(m_2+m'_2)^2$
(threshold of the channel 2), are explicitly written in (2.8) and
(2.9)  -- they are present in the phase space factors $\rho_1$ and
$\rho_2$, respectively.  The function $K(k^2)$ is real below the
threshold of the second channel, $(m_1+m'_1)^2<s<(m_2+m'_2)^2$,
because in this region  $\rho_2=i|\rho_2|$.

\subsection{Dispersion relation $N/D$-method and the  $K$-matrix
representation}

The dispersion relation $N/D$ method \cite{CM}
correctly reproduces analytic properties of the amplitude on the
whole $s$-plane. Following \cite{deut,as}, we present here the
elements of this method which are used in the analysis
of meson spectra.

Partial amplitude $A(s)$ (as before, the index $\ell$ is omitted for
 brevity) is written in form of the  ratio
$$A(s)=\frac{N(s)}{D(s)}\,.\eqno{(2.10)}$$
$N(s)$ is a function of the complex variable $s$. It has only left-hand
side  singularities of the
amplitude, which are related to the interaction
forces, i.e. to the diagrams with  meson exchanges in the crossing
channels (see Fig. 1). These singularities are
located on the left from the
threshold singularities, at  $s=(m_1+m'_1)^2-m^2_{crossing}$.

The $D$-function contains only right-hand side singularities, which
result from the rescattering of particles in the $s$-channel. Figure
2.1 shows the corresponding rescattering processes.

First, we consider the one-channel case, with equal
particle masses $m_1=m'_1$, in this case the $D$-function assumes
the following form:
$$
D(s)=1-B(s)\,,\qquad B(s)=\int_{4m^2}^{\infty}\frac{ds'}{\pi}
\frac{N(s')\rho(s')}{s'-s-i0}\,.
\eqno{(2.11)}
$$
Here index 1 is omitted:
$m_1\rightarrow m$, $\rho_1\rightarrow\rho$. The form of equation
(2.11) suggests that $D(s)\rightarrow 1$ at $s\rightarrow\infty$
(more generally, $D(s)\rightarrow Const$ at $s\rightarrow\infty$, since
this case can be reduced to Eq. (2.11) by the re-definition of $N(s)$).
Moreover, in Eq. (2.11) it is also suggested that the $D$-function
does not contain the Castillejo--Dalitz--Dyson poles (a detailed
description of the $N/D$-method may be found in Refs.
\cite{CM,deut}).

Representation of the $N$-function in the form of a sum of the
separable vertex functions \cite{deut} is likely to be a reasonable
anzatz: this technique is successfully applied for the description
of the $NN$ scattering amplitude \cite{as}; in addition, the technique
is elaborated for the presentation of the $t$-channel exchange diagrams
as a sum of separable vertex functions \cite{ks}. For the simplest
case, which is discussed below, $N(s)=g^2(s)$. Then
$$
A(s)=\frac{g^2(s)}{1-B(s)}\,,\qquad
B(s)=\int_{4m^2}^{\infty}\frac{ds'}{\pi}\frac{g(s')\rho(s')g(s')}
{s'-s-i0}\,.\eqno{(2.12)}$$
Expanding Eq. (2.12) in a series with respect to $B(s)$,
we represent the
amplitude $A(s)$ as a sum of diagrams shown in Figs. 2a, 2b, 2c,
etc.: $B(s)$ in Eq. (2.12) is a loop diagram.
 At $s>4m^2$, the loop diagram is a complex quantity:
$$ ImB(S)=g^2(s)\rho(s)\, , \qquad Re
B(s)=P\int_{4m^2}^{\infty}\frac{d(s')}{\pi}
\frac{g^2(s')\rho(s')}{s'-s}\,.
\eqno{(2.13)}
$$

The amplitude (2.12)  stands for the case when the partial wave
amplitude does
not contain  input particles: the bound states, if any, are formed
by particle interaction taken in the  $N$-function.
The inclusion of input particles
into the amplitude corresponds to the assumption on
$D(s)$ increasing  at $s\rightarrow\infty$.
The linearly growing
 $D(s)$  can be written in the form:
$$D(s)=m_0^2-s-B(s)\,,\qquad B(s)=\int_{4m^2}^{\infty}\frac{d(s')}{\pi}
\frac{g^2(s')\rho(s')}{s'-s-i0}\,.\eqno{(2.14)}$$
The amplitude
$$A(s)=\frac{g^2(s)}{m_0^2-s-B(s)}\eqno{(2.15)}$$
is an infinite set of diagrams shown in Figs. 2d, 2e, 2f,
etc.; $B(s)$ stands for the loop diagram and $(m_0^2-s)^{-1}$
is the propagator of the input particle.

The $K$-matrix representation of the $A(s)$ amplitude is
related to the explicit separation of the imaginary part
of the loop diagram:
$$
A(s)=\frac{g^2(s)}{m_0^2-s-ReB(s)-i\rho(s)g^2(s)}=\frac{K(s)}
{1-i\rho(s)K(s)}\, ,           \qquad
K(s)=\frac{g^2(s)}{m_0^2-s-ReB(s)}\, .
\eqno{(2.16)}
$$
In the two--particle loop diagram, the function $ReB(s)$ is
analytical at the point $s=4m^2$. This means that
poles are the only singulariries of $K(s)$ in the physical region.
However, at the left half-plane $s$, $K(s)$ has singulariries related
to the  $t$-channel exchanges.

The poles of the amplitude $A(s)$, which are determined by
the condition
$$
m_0^2-s-B(s)=0\, ,
\eqno{(2.17)}
$$
are related to particles with quantum numbers of the partial
wave under consideration. If
the pole is above the threshold, at $Re\,s>4m^2$,  we deal with
the resonance;  this very case is studied further. Let the equality
(2.17) be satisfied in the point:
$$
s=M^2\equiv\mu^2-i\Gamma\mu\,.  \eqno{(2.18)} $$
Expanding the real part of the denominator  (2.15) in a series near
$s=\mu^2$, one has:
$$ m_0^2-s-ReB(s)\simeq(1+Re B'(\mu^2))
(\mu^2-s)-ig^2(s)\rho(s)\,.  \eqno{(2.19)} $$
The standard Breit--Wigner approximation comes when $Im\,B(s)$
is fixed in the point $s=\mu^2$. If the pole is located not far from
the threshold singularity $s=4m^2$,  it is necessary to keep
the $s$-dependence in the phase space factor, and we use the
modified Breit--Wigner formula:
$$
A(s)=\frac{\gamma}{\mu^2-s-i\gamma\rho(s)}\, , \qquad
\gamma=\frac{g^2(\mu^2)} {1+Re B'(\mu^2)}\, .  \eqno{(2.20)} $$
Similar resonance approximation may be also carried out  for the
$K$-matrix amplitude representation, that corresponds to the expansion
for $K(s)$ given by Eq. (2.16)  near the point $s=\mu^2$:
$$K(s)=\frac{g^2(K)}{\mu^2-s}+f\,.\eqno{(2.21)}$$
Here
$$
g^2(K)=\frac{g^2(\mu^2)}{1+Re B'(\mu^2)}\,, \qquad
f=\frac{g^2(\mu^2)}{2(1+Re B'(\mu^2))}-\frac{2g(\mu^2)g'(\mu^2)}
{1+Re B'(\mu^2)}\,.
\eqno{(2.22)}
$$

\subsection{Multichannel scattering}

The resonance amplitude (2.15) can be
easily generalized for the $n$ channel
case. The corresponding
transition amplitude $b\rightarrow a$ is equal to:
$$
A_{ab}(s)=\frac{g_a(s)g_b(s)}{m_0^2-s-B(s)}\,,\qquad
B(s)=\sum_{c=1}^{n}B_{cc}(s)\,,
\eqno{(2.23)}
$$
where $B_{cc}$ is defined by the standard expression (see Eq. (2.14)),
with the properly chosen phase space factor, vertex function
and integration region:
$$g^2(s')\rho(s')\rightarrow g_c^2(s')\rho_c(s')\, , \qquad 4m^2
\rightarrow 4m^2_c \eqno{(2.24)} $$
The transition amplitudes $A_{ab}$ form a matrix $\hat A$. The
 amplitude written in the $K$-matrix representation reads:
$$ \hat A=\hat
K\frac{I}{I-i\hat \rho\hat K}\,, \eqno{(2.25)} $$
where $\hat K$ is the $n\times n$ matrix,  $K_{ab}(s)=K_{ba}(s)$,
$I$ is the unit $n\times n$ matrix $I=diag(1,1,\ldots,1)$, and $\hat
\rho$ is the diagonal matrix of phase space factors:
$$ \hat
\rho=diag(\rho_1(s),\rho_2(s),\ldots,\rho_n(s))\,.  \eqno{(2.26)} $$
The $K$-matrix elements are equal to:
$$
K_{ab}(s)=\frac{g_a(s)g_b(s)}{m_0^2-s-Re B(s)}\,.
\eqno{(2.27)}
$$
In the vicinity of the resonance, the $K$-matrix elements may be
expanded in a series:  in this case we have
a representation of the $K$-matrix elements
similar to that of Eq. (2.21).

\subsection{$q\bar q$-mesons:
the problem of small and large distances}

The $q\bar q$ classification of
meson states in the vicinity of 1000--2000
MeV faces the problem of the quark--hadron duality as well as
a tightly related problem of separating  large- and small-$r$
interactions contributing
into formation of meson spectra.

Let us discuss these problems using the language of the standard quark
model. In this model the $q\bar q$ levels are determined by the
potential which increases infinitely with $r$:
$V(r)\sim\alpha r$ (see Fig. 3a). Infinitely rising potential
creates an infinite set of the  $q\bar q$ levels. However, it is
obvious that the standard quark model is a
simplified picture, since only the
lowest $q\bar q$-levels are stable with respect to hadron decays. The
heavier states are decaying by hadron channels: an excited $(q\bar
q)_a$-state produces new $q\bar q$ pair, then the $(q\bar q)_a+(q\bar
q)$ quarks recombine into mesons, which leave the confinement trap
with a formation of the continuous meson spectrum.  This structure is
conventionally shown in Fig. 3b, where the interaction related to the
confinement is shown as a certain potential barrier, namely, the
interaction at  $r<R_{confinement}$  creates the discrete  levels of
$q\bar q$ spectra, while the transitions into the region
$r>R_{confinement}$ give rise to the continuous meson spectrum.  This
very meson spectrum is observed in the experiment. This problem of
reconstruction of the $q\bar q$-levels
created at $r<R_{confinement}$ is
directly related to the determination of the effect of the meson
decay spectra on the level shift:
classification of
 $\#$ levels requires elimination of the components of real mesons
that are the decay products of real meson.

The $K$-matrix representation of the amplitude resolves the problem
of excluding the  components of real  mesons;
formally, it corresponds to the limit $\rho_a\rightarrow 0$ in
Eq. (2.25). If only leading pole singularities are taken into
account, the transition amplitude  $b \to a$ assumes the form:
$$
A_{ab}^{\rm bare}(s)= K_{ab}(s)=\frac{g_a(K)g_b(K)}{\mu^2-s}+f_{ab}\,.
\eqno{(2.28)} $$
Thus, the $K$-matrix pole corresponds to the state with the
removed cloud
of real mesons. For this reason, we call corresponding
states  "bare mesons" \cite{km1500,km1900}. However, one should
distinguish between this notation and that of "bare particles" of
field theory, where a cloud includes  virtual mass-off-shell particles
as well.

In the case when the $q\bar q$-spectrum contains several states with the
same quantum numbers, the amplitude $A_{ab}^{\rm bare}(s)$  is
determined by the sum of the corresponging poles:
 $$ A_{ab}^{\rm bare}(s)=\sum_{\alpha}\frac{g_a^{(\alpha)}(K)
g_b^{(\alpha)}(K)}{\mu_\alpha^2-s}
+f_{ab} \,.
\eqno{(2.29)}
$$

Representation of the amplitude
responsible for the interaction at $r<R_{confinement}$
in the form of a series of
poles is not new: it was widely used in dual models for
contributions which are leading in the $1/N_c$-expansion. From the
point of view of these models, the $s$-independent term $f_{ab}$
is the thorough contribution of poles which are distant from the region
under consideration.

Coupling constants of the bare state, $g_a^{(\alpha)}(K)$, are a source
of information on the quark--gluon content of this state.

\subsection{Coupling constants of the gluonium and $q\bar q$-states to
meson channels: the rules of the $1/N$-expansion and quark combinatoric
relations}

Quark--gluon content of states related to the $K$-matrix poles (bare
states) is revealed in relations between couplings of
these poles to meson channels, $g_a^{(\alpha)}$.

First, let us evaluate these couplings using the rules of
$1/N$-expansion; this evaluation will be done both for the
transitions {\bf glueball} $\rightarrow$ {\bf two mesons} and for the
transitions $q\bar q-${\bf state} $\rightarrow$ {\bf two mesons}.
For this purpose, we
consider the gluon loop diagram which corresponds to
the two--gluon self--energy part: {\bf glueball}
$\rightarrow$ {\bf two gluons} $\rightarrow$ {\bf glueball} (see Fig.
2.4a). This loop diagram is of the order of unity, provided the
glueball is a two--gluon composite particle:
$ B(G\rightarrow gg\rightarrow G)\sim
g_{G\rightarrow gg}^2N_c^2\sim 1$, where $g_{G\rightarrow gg}$ is
a coupling constant of a glueball to two gluons.  Therefore,
$$
g_{G\rightarrow gg}\sim 1/N_c\,.\eqno{(2.30)} $$
Coupling constant for the transition  $g_{G\rightarrow q\bar q}$
is determined by the diagrams of Fig. 4b. Similar evaluation gives:
$$ g_{G\rightarrow q\bar q}\sim
g_{G\rightarrow gg}g_{QCD}^2N_c\sim 1/N_c\,.\eqno{(2.31)} $$
Here $g_{QCD}$ is the quark--gluon coupling constant, which is of the
order of $1/\sqrt{N_c}$ \cite{thooft}.
Coupling constant for the transition {\bf glueball} $\rightarrow$ {\bf
two mesons} in the leading $1/N_c$ terms is governed by the
diagrams of Fig. 4c:
$$ g_{G\rightarrow mm}^L\sim
g_{G\rightarrow q\bar q}g_{m\rightarrow q\bar q}^2N_c\sim
1/N_c\,.\eqno{(2.32)} $$
Here the approximate equality $g_{m\rightarrow q\bar q} \sim
1/\sqrt{N_c}$ is used which follows from the fact that the
loop diagram of the meson propagator  (see  Fig. 4d) is of the order
of the unity: $B(m\rightarrow q\bar q\rightarrow m)\sim
g_{m\rightarrow q\bar q}^2 N_c\sim 1\,.$ The diagram of  Fig.
2.4e-type  governs the couplings for the transition {\bf glueball}
$\rightarrow$ {\bf two mesons} in the next-to-leading terms of the
$1/N_c$-expansion:
$$ g_{G\rightarrow mm}^{NL}\sim g_{G\rightarrow
gg}g_{QCD}^4g_{m\rightarrow q\bar q}^2N_c^2 \sim 1/N_c^2\,.
\eqno{(2.33)} $$

As was mentioned above, the glueball can decay into channels (1.8). In
this way, the production of light quarks by gluons is going on with a
violation of the flavour blindness, $u\bar u:
d\bar d:  s\bar s=1:1:\lambda$. Within such an assumption, the coupling
constants for the transition {\bf glueball} $\rightarrow$ {\bf two
pseudoscalar mesons} can be calculated using quark combinatorics. The
rules of quark combinatorics were successfully applied to the
calculation of yields of secondary particles in the hadron--hadron high
energy collisions \cite{ash} and to the decay
$J/\psi\rightarrow${\bf hadrons} \cite{w}.  The calculation of
couplings in the decays {\bf glueball} $\rightarrow$ {\bf mesons}
has been carried out in Refs.  \cite{ac,a95,glp}.

The glueball  decay couplings  to channels (1.8) are given in
Table I for  leading, $g_{G\rightarrow mm}^L$,
and next-to-leading, $g_{G\rightarrow mm}^{NL}$, terms
of the $1/N$-expansion.  Unknown dynamics of the decay is hidden
in the parameters $G_L$ and $G_{NL}$.  The decay constant to the
 channel $n$ is a sum of both contributions:
$$
g_{G\rightarrow mm}^L(n) +g_{G\rightarrow mm}^{NL}(n)\,.  \eqno{(2.34)}
$$
The second term is suppressed, as compared to the first one, by the
factor $N_c$; the experience in calculation of quark diagrams
teaches us that this suppression is of the order of 1/10.

The sum of couplings squared satisfies the sum rules:
$$
\begin{array}{l}
\sum_{n}\left(g_{G\rightarrow mm}^L(n)\right)^2I(n)=\frac 12 G_L^2
(2+\lambda)^2\,, \qquad
\sum_{n}\left(g_{G\rightarrow mm}^{NL}(n)\right)^2I(n)=\frac 12 G_{NL}^2
(2+\lambda)^2\,,\end{array}\eqno{(2.35)}
$$
where $I(n)$ is an identity factor for the particles produced, see
Table I. These sum rules follow from the quark--hadron duality:
the sum of squared coupling constants over the whole set of flavour
states is equivalent to the sum of cut diagrams with the quark loops
(diagrams of the type of Fig. 4f for leading and of the type of
Fig.  4g for next-to-leading terms). Each quark loop contains the
factor $(2+\lambda)$ related to the summation over light
flavours, see Eq. (1.10).

Quark combinatoric rules may be applied to the calculation of couplings
 of  $(q\bar q)_a$-mesons to pseudoscalar channels  (1.8).  There
exist two types of transitions $(q\bar q)_a-${\bf
state}$\rightarrow${\bf two mesons}: they are shown in Fig. 5a and
2.5b.  The type of process
represented by the diagram of Fig. 5a
is leading according to the rules of $1/N$ expansion; its coupling
constant is of the order of
$$ g_{m(a)\rightarrow
mm}^L\sim g_{m\rightarrow q\bar q}^3 N_c\sim \frac{1}{\sqrt{N_c}}\,.
\eqno{(2.36)} $$
The decay constant for the process of Fig. 5b  is of the order of
$$
g_{m(a)\rightarrow mm}^{NL}\sim g_{m\rightarrow q\bar q}^3N_c^2
g_{m\rightarrow q\bar q}^4\sim\frac{1}{N_c\sqrt{N_c}}\,.  \eqno{(2.37)}
$$
The coupling constants for the decays $(q\bar q)_a \rightarrow\pi\pi$,
$K\bar K$, $\eta\eta$, $\eta\eta'$, $\eta'\eta'$ are shown in Table 2
both for the leading and next-to-leading orders; $g^L$ and $g^{NL}$ are
parameters, in which the unknown dynamics of soft decay is hidden.
Concerning
 the glueball decay, the coupling for the $(q\bar q)_a$-meson
decay to the channel  $n$ is the sum of both terms:
 $$ g_{m(a)\rightarrow
mm}^L(n)+g_{m(a)\rightarrow mm}^{NL}(n)\,.  \eqno{(2.38)} $$
These two terms in Eq. (2.38) define the decay constant of the $(q\bar
q)_a$-meson in general case: different variants of fixing the ratios of
coupling constants correspond to different choices of the ratio
 $g_L/g_{NL}$. The examples of different fixations of
$g_L/g_{NL}$ may be found in \cite{bonn,glp}.

Let us stress once again that the coupling constant ratios for
the $(q\bar q)_a$-states (Table II) become identical to
those of the glueball,
 when  $\phi=\arctan\sqrt{\frac{\lambda}{2}}$:
it is valid for leading and next-to-leading contributions. Therefore,
that, on the basis of a study of couplings to the hadron decay channels
only, it is impossible to distinguish between the
glueball and the $I=0$ $(q\bar q)_a$-meson with the mixing angle $\phi
$ close to $30^o$.

\newpage
\centerline{\hspace{-6cm}\epsfig{file=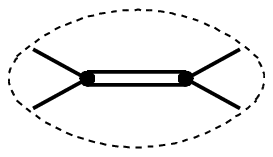,width=4.5cm}}

\vspace{-3.8cm}

\centerline{\epsfig{file=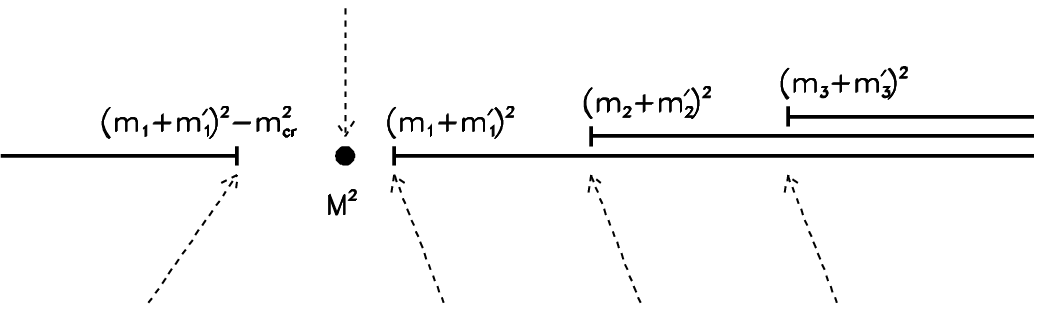,width=18cm}}

\vspace{-1.6cm}

\centerline{\hspace{0cm}\epsfig{file=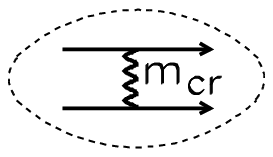,width=4.5cm}
            \hspace{0cm}\epsfig{file=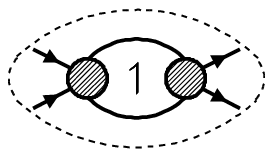,width=4.5cm}
            \hspace{-0.75cm}\epsfig{file=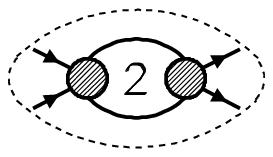,width=4.5cm}
            \hspace{-1cm}\epsfig{file=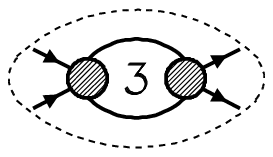,width=4.5cm}}
Fig. 1. Complex  $s$-plane and positions of singularities of the
 partial amplitude:  right-hand  singularities at $s\ge (m_1+m'_1)^2$
are due to the elastic and inelastic rescatterings,  left-hand  ones
are due to the interaction forces, that is, particle exchanges in the
crossing channels.

\centerline{    \epsfig{file=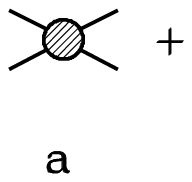,height=5cm}
\hspace{-3.3cm} \epsfig{file=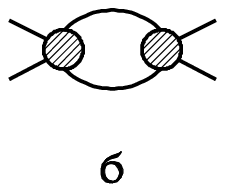,height=5cm}
\hspace{-1.7cm} \epsfig{file=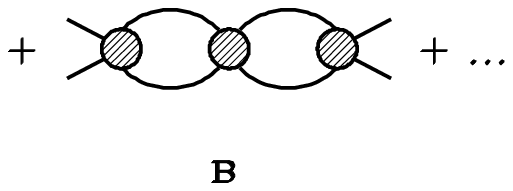,height=5cm}}

\vspace{-2cm}

\centerline{   \epsfig{file=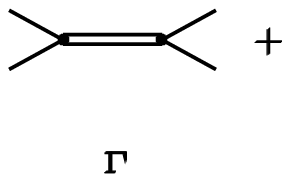,height=4cm}
\hspace{-1.3cm}\epsfig{file=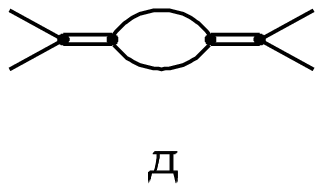,height=4cm}
\hspace{-1.2cm}\epsfig{file=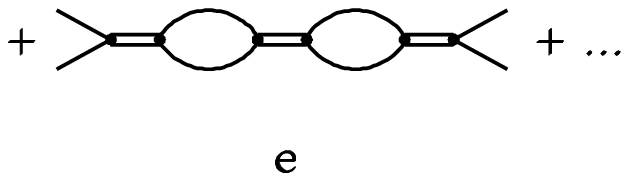,height=4cm}}
Fig. 2. Diagrams standing for the  $s$-channel scattering.

\centerline{\epsfxsize=11cm \epsfbox{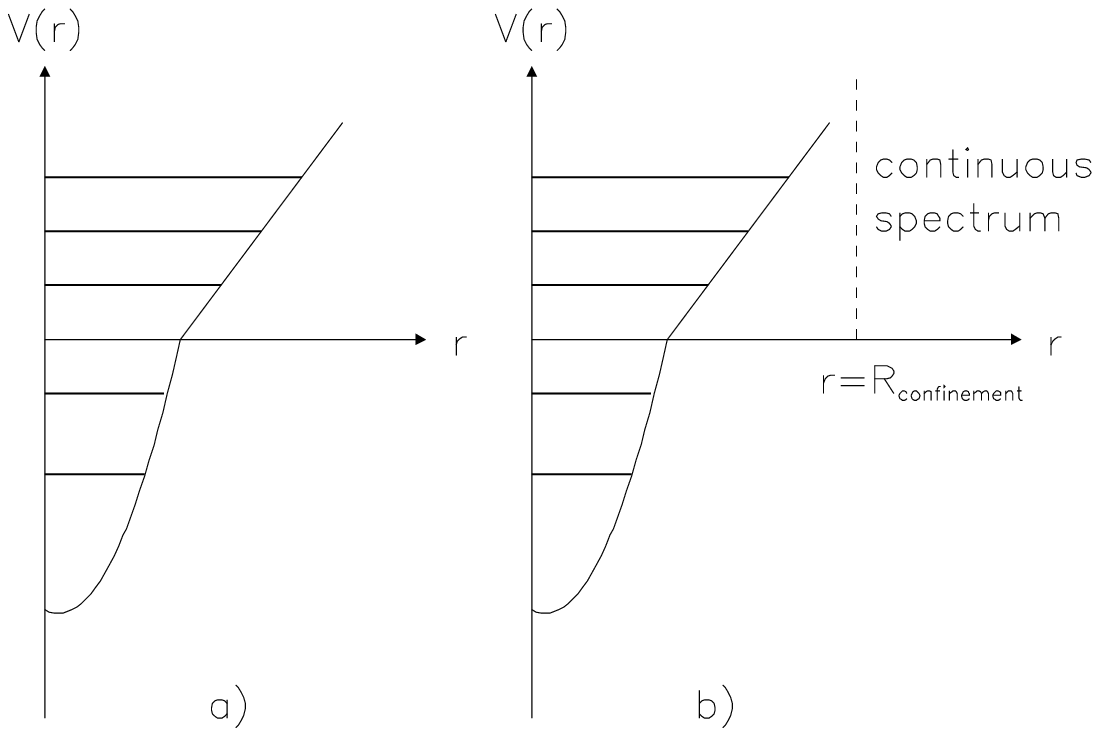}}
Fig. 3. a) Standard
quark model potential with stable $q\bar q$-levels,
b) Potential with the unstable highly excited levels which corresponds
to the realistic situation for  $q\bar q$ states.

\centerline{\epsfxsize=11cm \epsfbox{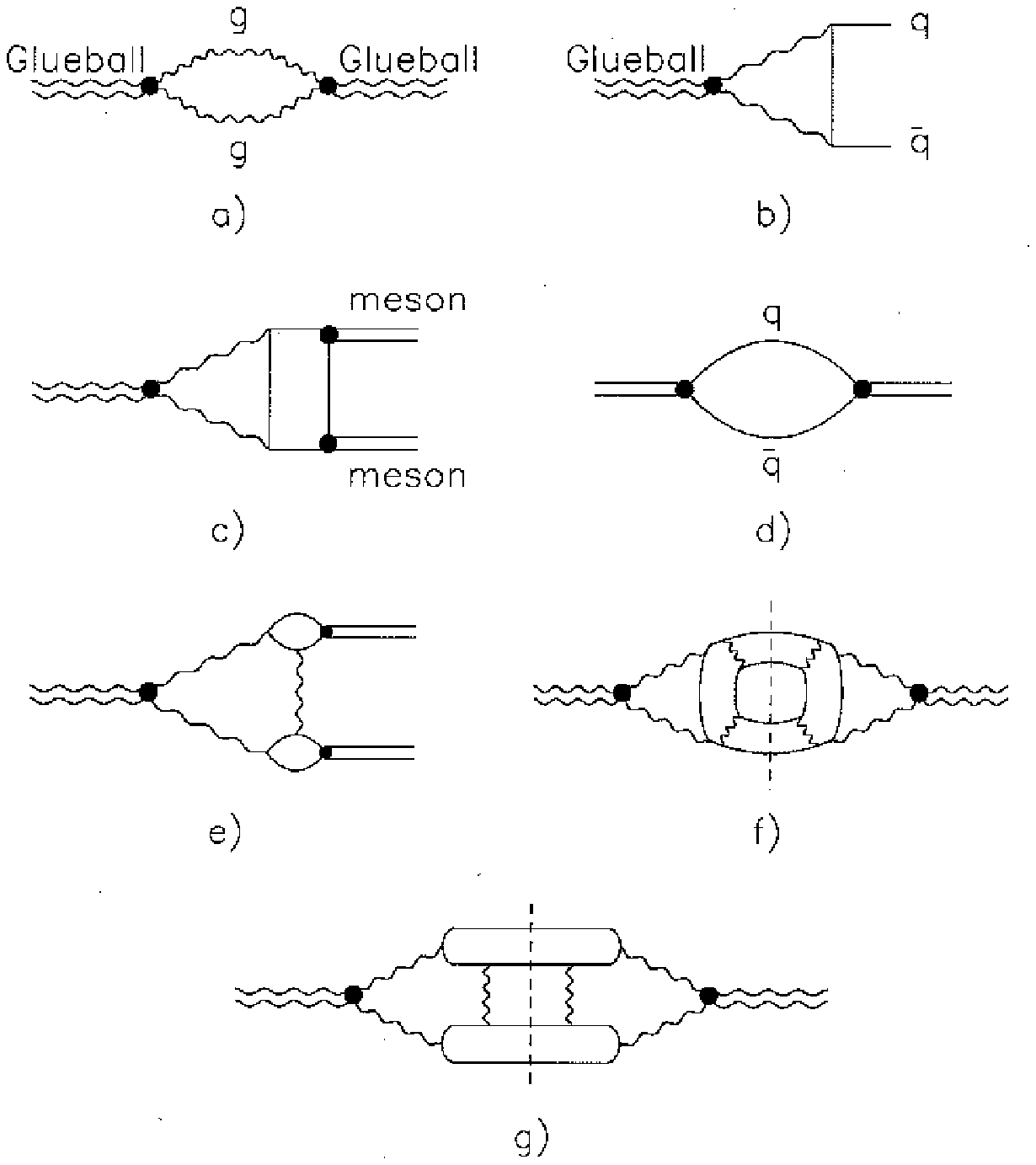}}
Fig. 4. Diagrams for the glueball decay into two mesons.

\centerline{\epsfxsize=11cm \epsfbox{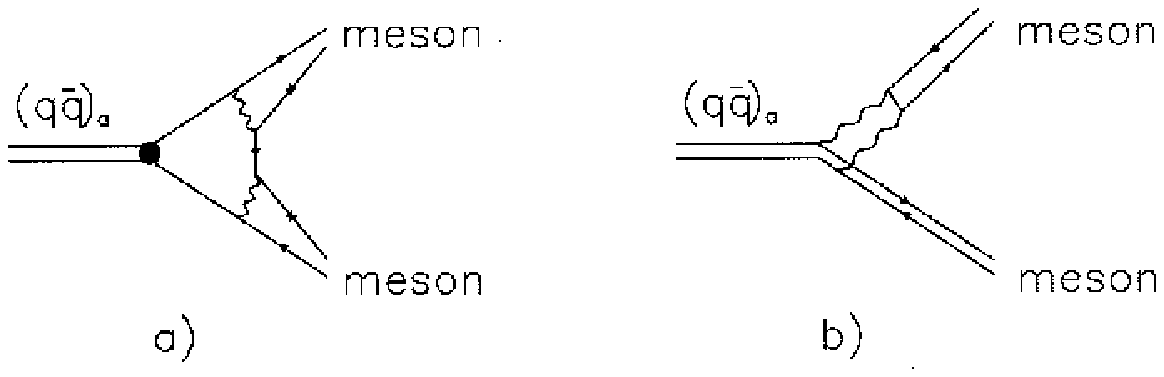}}
Fig. 5. Diagrams for the decay of the  $(q\bar q)_a$-state into two
mesons.

\section{ $K$-matrix analysis of meson spectra and  nonet classification
 of $q\bar q$-states }

In this Section, the results obtained in Ref.
\cite{km1900,fullkm,kmkpi} for the waves $00^{++}$, $10^{++}$
$02^{++}$, $12^{++}$ and $\frac12 0^+$ are presented. On the basis of
this analysis, the nonet classification of  $q\bar q$-states is
established.

\subsection{ $K$-matrix fit of $00^{++}$-wave: the spectra
$\pi\pi$, $K\bar K$, $\eta\eta$ and $\eta\eta'$}

To describe the spectra, in Ref. \cite{km1900} the standard
$K$-matrix representation of the  $00^{++}$ amplitude (2.25) was used,
where $K_{ab}$ is the $5\times 5$ matrix $(a, b=1,2,\ldots,5)$, with
the following notations for channels: $1=\pi\pi$, $2=K\bar K$,
$3=\eta\eta$, $4=\eta\eta'$, 5=($\pi\pi\pi\pi+${\it other multimeson
states}).

Matrix elements of  $K_{ab}$ are parametrised in the form similar to
Eq.  (2.28):
$$ K_{ab}=\left ( \sum_\alpha \frac{g^{(\alpha)}_a
g^{(\alpha)}_b}
{M^2_\alpha-s}+f_{ab}\frac{1 \, \mbox {GeV}^2+s_0}{s+s_0} \right ),
\eqno{(3.1)}$$
with the restriction $s_0 \geq 1 \; GeV^2$.

The following
formulae provide the description of the spectra
 $\pi\pi$, $\eta\eta$ and $\eta\eta'$, obtained by GAMS  for
the reactions with the $t$-channel pion exchange:
$$A_{\pi N\rightarrow Nb}=
g(\bar \psi_N\gamma_5\psi_N)F(t)D(t)\;
 K_{1a}(t) (1-i\hat \rho  \hat K)^{-1}_{ab},
\eqno{(3.2})$$
$$
K_{1a}(t)=\left(\Sigma_{\alpha}\frac{g_1^{(\alpha)}
(t)g_a^{(\alpha)}}{M_{\alpha}^2-s}+f_{1a}(t)\frac{1 \,\mbox
{GeV}^2+s_0}{s+s_0} \right).
\eqno{(3.3)}
$$
Here $D(t)$ is the pion propagator,  $F_N(t)$ is
the nucleon form factor
related to the vertex $\pi NN$, and $g_1^{(\alpha)}(t)$ and $f_{1a}(t)$
are  form factors of the  pion block.

The piece of the amplitude for the reaction
$p\bar p\; (at\;\;rest)\rightarrow
\pi^\circ\pi^\circ\pi^\circ$, $\pi^\circ\eta\eta$, which describes
the production of two mesons in the $00^{++}$ state, is written as
follows:
 $$ A_{p\bar p \; \rightarrow{\rm mesons}}=
A_1(s_{23})+A_2(s_{13})+A_3(s_{12})
\eqno{(3.4)}
$$
The amplitude  $A_k(s_{ij})$ corresponds to the process with the  "last
interaction" of particles $ij$, while  the particle
$k$ remains a spectator.

The amplitude $A_1(s_{23})$ for the spectra $\pi\pi$ and $\eta\eta$
has the following form ($b=\pi\pi$, $\eta\eta$):
$$ A_1(s_{23})=
K_{p\bar p\pi,a}
(s_{23})\left(1-i \rho K\right)^{-1}_{ab}, \;
 K_{p\bar p\pi,a}(s_{ij})=\left (
\sum_\alpha \frac{\Lambda^{(\alpha)}_{p\bar p\pi} g^{(\alpha)}_a}
{M^2_\alpha-s_{ij}}+\phi_{p\bar p\pi,a}\;
\frac{1\,\mbox{GeV}^2+s_0}{s_{ij}+s_0} \right)\; .
\eqno{(3.5)}$$
In the reaction
$p\bar p\; (at\;rest )\rightarrow \pi^\circ\pi^\circ\pi^\circ$,
the amplitude is symmetric with respect to the permutation of pion
indices: $A_1(s_{ij})=A_2(s_{ij})=A_3(s_{ij})$.  The
$\pi^\circ\pi^\circ$ interaction in the reaction $p\bar p\; (at\; rest)
\rightarrow \pi^\circ\pi^\circ\eta$ is determined as follows
(below $b=\pi\pi$):
$$ A_1(s_{23})= K_{p\bar p\eta,a} (s_{23})\left(1-i \rho
K\right)^{-1}_{ab}\; ,\qquad K_{p\bar p\eta,a}(s_{ij})=\left (
\sum_\alpha \frac{\Lambda^{(\alpha)}_{p\bar p\eta} g^{(\alpha)}_a}
{M^2_\alpha-s_{ij}}+\phi_{p\bar p\eta,a}\;
\frac{1\,\mbox{GeV}^2+s_0}{s_{ij}+s_0} \right)\; .
\eqno{(3.6)}$$
Parameters $\Lambda_{p\bar p\pi}^\alpha$, $\phi_{p\bar p\pi}$,
$\Lambda_{p\bar p\eta}$ and $\phi_{p\bar p\eta}$ can be complex
quantities with different phases, that follows from  the three--particle
interaction; more detailed discussion of the amplitude
$p\bar p\rightarrow$ {\bf three mesons} is given in
Ref. \cite{3body}.

\subsection{Results of the $K$-matrix fit for the $00^{++}$ wave in the
region below 1900 MeV }

Simultaneous $K$-matrix fit of the  $00^{++}$ spectra in the mass
region 550--1900 MeV performed in \cite{km1900} pointed out to the
existence of five bare states,  $f_0^{\rm bare}$. Only two of them,
 $f_0^{\rm bare}$(720) and $f_0^{\rm bare}$(1810),
contain a large $s\bar s$-component. This means that only two
$^3P_0q\bar q$-nonets can be built in the mass region below  1900 MeV.

The following requirements provide the ground for the nonet
classification of bare states:\\
(1) Nonet partners are orthogonal in
the flavour space, i.e. they must have the mixing angle differences
(see Eq. (1.9)) equal to $90^\circ$:  $\phi_1-\phi_2=90^\circ$
(for this value the corridor  $90^\circ\pm 5^\circ$ is allowed).\\
(2) Coupling constants $g^L$ and $g^{NL}$ (see Table 2) are
approximately equal for nonet partners:  $g_1^L\simeq
g_2^L$ and $g_1^{NL}\simeq g_2^{NL}$.

The standard quark model requires the equality of coupling
constants. However, the $s$-dependent vertex functions and loop
diagrams violate this equality because of the presence of mass
differences for nonet partners.  Moreover, the $K$-matrix coupling
constants have an additional $s$-dependent factor $(1+B'(s))^{-1}$ (see
Eq. (2.22)): this factor strongly affects the region of small masses
 (the region of the basic nonet $1^3P_0$), where the thresholds and
left-hand singularities of partial amplitude play more significant
role.

The fitting to experimental data (1.4)-(1.7) resulted in two
solutions, I and II. First, let us sum up the results for Solution I:
 $$ \begin{array}{lllll} \multicolumn{2}{c}{\mbox{Type of the state}}
 & \multicolumn{3}{c}{\mbox{Flavour wave function}}\\
 f_0^{bare}(720)& \to 1^3P_0q\bar q &
 0.40n\bar n-0.92s\bar s \\
 f_0^{bare}(1260)&\to 1^3P_0q\bar q & 0.92n\bar n+0.40s\bar s
 \\
 f_0^{bare}(1600)&\to 2^3P_0q\bar q & 0.995n\bar n-0.10s\bar s
 \\
 f_0^{bare}(1810)&\to 2^3P_0q\bar q & 0.10n\bar n+0.995s\bar s
 \\
 f_0^{bare}(1235)&\to  Glueball \to & 0.91n\bar n+0.42s\bar
 s. \\
\end{array}
\label{flav1}
\eqno{(3.7)}
$$
In Eq. (3.7) the  "flavour wave function" is
 introduced for the glueball:  it describes the  flavour
content of intermediate state for the glueball decay: see Fig. 4c.

Now let us summarize the results for Solution II:
$$
 \begin{array}{lllll}
 \multicolumn{2}{c}{\mbox{Type of the state }} &
 \multicolumn{3}{c}{\mbox{Flavour wave function }}\\
  f_0^{bare}(720)& \to 1^3P_0q\bar q  & 0.40n\bar n-0.92s\bar s
  \\
  f_0^{bare}(1260)&\to 1^3P_0q\bar q  &
  0.92n\bar n+0.40s\bar s \\
  f_0^{bare}(1235)&\to 2^3P_0q\bar q  & 0.74n\bar n+0.67s\bar s
  \\
  f_0^{bare}(1810)&\to 2^3P_0q\bar q  &
  0.67n\bar n-0.74s\bar s \\
  f_0^{bare}(1600)&\to  Glueball \to  & 0.91n\bar n+0.42s\bar s.
  \\
\end{array}
\eqno{(3.8)}
$$
The quality of data description by Solutions I and II can be seen in
Figs.  6-3.4 (dashed and solid curves, respectively).

\subsection{The resonances: are they bumps or dips in the spectra?}

During decades the search for meson resonances meant the search for
 bumps in the particle spectra. Only recently the understanding came
that it is not always so, and the resonance $f_0(980)$ provided us
with an example.  In the peripheric $\pi\pi$-spectra it reveals itself
as a dip, and a set of papers was devoted to this phenomenon, see
\cite{pdg,buggzou}.  The study of $00^{++}$ wave proved that meson
resonances in the region 1000-1600 MeV appear not only as bumps or
dips, but as shoulders in the spectra as well. The main particular
feature of the resonance is not a bump or a dip in the spectrum,
but a circle on the Argand diagram.

Figures 10 and 11 demonstrate the Argand diagrams relevant to the
fits of spectra under discussion. In Fig. 10 one can see the
$00^{++}$ amplitudes $A_{\pi\pi\rightarrow\pi\pi}$,
$A_{\pi\pi\rightarrow K\bar K}$, $A_{\pi\pi\rightarrow\eta\eta}$ and
$A_{\pi\pi\rightarrow\eta\eta'}$ as functions of the energy. Indeed,
 there exist rather legible circles obtained for the resonances
$f_0(980)$, $f_0(1300)$, $f_0(1500)$ and $f_0(1780)$. The manifestation
of resonances  $f_0(980)$ and $f_0(1300)$ in the form of circles is
rather clear for the amplitudes $A_{\pi(t)\pi\rightarrow\pi\pi}$
at large  $|t|$,  see  Fig. 11.

\subsection{Resonance $f_0(980)$: is it the $K\bar K$-molecule or the
descendant of the lightest scalar $q\bar q$-states?}

This is a principal problem for the  $q\bar q$-systematics that was
 firstly investigated  in \cite{12} by using the  $K$-matrix analysis
for the low--energy part of the $00^{++}$-wave. Following this paper,
we provide arguments that $f_0(980)$ is the descendant of the lightest
$q\bar q$-states.

Two poles correspond to the resonance $f_0(980)$, with the
following complex masses:  $M=1015-i46 $ MeV on the second sheet
(under the  $\pi\pi$ cut) and $M=936-i238$ MeV  on the third sheet
(under the $\pi\pi$ and $ K\bar K $ cuts).  The second pole, at
$936-i238$ MeV, appears because of the  well-known
doubling of poles affected by
  the proximity of the  $K\bar K$-threshold --  for
example, see Refs.  \cite{buggzou,flatte}.  The  first
pole at $1015-i46$ MeV dominates in the $\pi\pi$-spectrum, providing a
sharp dip in the  $\pi\pi\rightarrow\pi\pi$ spectrum or a bump in
the spectrum $\pi\pi(t)\rightarrow\pi\pi$ at large $|t|$ \cite{G1}.
Let us study the dynamics of this pole
 when the decays into the channels $1=\pi\pi$ and $2=K\bar K$
are subsequently switched on and  off.   To this end,
let us make the following replacement in the  $K$-matrix amplitude:
$$
g_{1}(720)\rightarrow\xi_1 g_{1}(720)\,,\qquad
g_{2}(720)\rightarrow\xi_2 g_{2}(720)\,,\ \eqno{(3.9)} $$
with the parameters
$\xi_a$ varying in the interval $0\leq \xi_a \leq 1\,$.  At
$\xi_1 \to 0$ and $\xi_2 \to 0$ the decay channels for the lightest
$00^{++}$ state are switched off, and we deal with the "bare" state, in
terms of Refs. \cite{km1500,km1900}. At $\xi_1=\xi_2=1$
the real case is restored.  At $\xi_1 \simeq \xi_2 \simeq 0$ the mass
of the bare state is in the vicinity of 720 MeV, while, with the
increase of $\xi_a$, the lightest scalar state acquires the components
of real mesons, $\pi\pi$ and $K\bar K$, and, due to  transitions
to these states, it mixes with the other scalar ones. As a result,
the mass of the lightest scalar state increases, approaching the region
around 1000 MeV. At $\xi_1=\xi_2=1$ the pole of the amplitude is at
$$
M(real\,\,position)=1015-i46\; \mbox{MeV} \,, \eqno{(3.10)}
$$ i.e.
near the  $K\bar K$-threshold. Therefore, the  $K\bar K$-component is
of the
quasi-molecular type: the relative momentum of $K$-masons is small,
hence the mean distance between the $K$-mesons is comparatively large.
However, one can see that the $K\bar K$-component weakly affects
the formation of the final state.  Indeed, let us switch off the $K\bar
K$-state, i.e. let us put in the amplitude $\xi_{1}=1$
and $\xi_{2}=0$.  Then, the pole appears in the  point:
$$
M(without\,\,K\bar K)=979-i53\; \mbox{MeV} \,.  \eqno{(3.11)} $$
The mass shift
$$ M(real \,\,position )-M(without
\,\,K\bar K)=36+i7\; \mbox{MeV} \eqno{(3.12)} $$
is actually not large,
thus proving  the role of the $K\bar K$-component in the formation
of the real $f_0(980)$ state be insignificant.

Concluding, the  $K$-matrix analysis of  $00^{++}$-wave restores
the following picture of the formation of $f_0(980)$.  Before
mixing, there existed the lightest scalar/isoscalar $q\bar q$-state
$f_0^{bare}(720\pm 100)$, with the flavour wave function close to the
octet one:  $$ \psi_{flavour}(720)=Cos\theta_S\, \psi_8 -Sin\theta_S
\,\psi_1\; ,\qquad \theta_S =14^o \pm 12^o \; , \eqno{(3.13)} $$
$$
\psi_1=\frac{1}{ \sqrt{3} }(u\bar u +d\bar d +s\bar s )\; \qquad
\psi_8=\frac{1}{ \sqrt{6} }(u\bar u +d\bar d)- \sqrt{ \frac{2}{3} }
s\bar s \; .  \eqno{(3.14)} $$
The mixing with other states, which is going through the transition
 $f_0^{bare}(720)\rightarrow\pi\pi$, leads to the formation of the
resonance with the characteristics which are almost the same as those
 observed at the experiment, see Eq. (3.11).  The onset of
the $K\bar K$-component $f_0^{bare}(720)\rightarrow K\bar K$ affects
 relatively small shift of  mass and width, see Eq. (3.12).

It should be noted that direct measurements also point out
that $f_0(980)$ has  considerable small-$r$
components: the production of  $f_0(980)$ is not suppressed in the
reaction $\pi^-p\rightarrow f_0(980)\; p$ at large momentum transfers
\cite{G1,gams} as well as in  radiative  $J/\psi$ decay
 \cite{china}.

One should pay attention to the fact that the lightest scalar/isoscalar
state, that is, $\eta $-meson, also has flavour wave function  close
to the octet one:  $\eta =Cos\theta_P \,\psi_8
-Sin\theta_P \,\psi_1 $ with $\theta_P =-16.7^o \pm 2.8^o $ \cite{11}.

\subsection{The wave $IJ^{PC}=10^{++}$ }

Two isovector/scalar resonances are clearly seen in the reactions (1.4)
\cite{15,16}. The lightest of them is a well-known  $a_0(980)$,
while the next resonance is the recently discovered
$a_0(1450)$:  according to PDG compilation \cite{pdg}, its mass is
$1450\pm 40$  MeV and the width is $\Gamma=270\pm 40$ MeV. Note
that the fitting to the latest high statistics data
\cite{km1900,fullkm,bugg} provided us with a  greater mass value --- it
is equal to $1520\pm 40$ MeV.

For the description of the scalar/isoscalar amplitude, in Ref.
\cite{fullkm} the two-pole $4\times 4$ matrix was used, with the
channel notations $1=\pi\eta ,\; 2=K\bar K ,\; 3=\pi\eta' , \;
4=multiparticle\,\, states$. The couplings to the two-meson channels
are defined by the quark combinatoric relations, see Table 3: we remind
that the constants $g^L$  are the same for all multiplet members.
At the first stage of the fit, the coupling constants of the lightest
resonance $a_0$ have been varied in the interval limited by the
constants $g^L[f^{bare}_0(720)]$ and $g^L[f^{bare}_0(1260)]$.
In all the variants of the fit, the coupling
$g^L[a^{bare}_0(lightest \; state)]$ was found to be rather close
to $g^L[f^{bare}_0 (1260)]$: in the final variant of the fit
these coupling
constants are taken equal to each other. The couplings of the
next isovector/scalar resonance to two mesons are also fixed:  they
are equal to each other for all the multiplet  members $2^3P_0$.

The fit allowed us to find two solutions for the wave  $10^{++}$,
which practically coincide for the  resonance
sector and differ for  background terms. Positions of the
amplitude poles and relevant bare states are shown in Eqs. (1.14) and
(1.15).

In Ref. \cite{minion}, the hypothesis was  discussed about
resonances  $a_0(980)$ and $f_0(980)$ belonging to a special class
of states (minions), which  are loosely bound to hadron channels: the
small widths of $a_0(980)$ and $f_0(980)$ were considered as
arguments in favour of that particular nature.
 The characteristics of
 $a_0(980)$ provide a good opportunity to check this
hypothesis, because actually the components of real mesons yielded by
the decays $a_0(980) \to \pi\eta\,,K\bar K$ do not influence this
state, see Eqs. (1.14) and (1.15).  However, the $K$-matrix
 data fittings \cite{fullkm} proved that $g^L[f^{bare}_0 (964)]$  is
 not small, being of the order of a standard hadronic value, for the
 small width of $a_0(980)$ is related not to the probability of the
 decay, as it could follow from the  minion nature of $a_0(980)$, but
 to the threshold effect. It should be emphasized  that this result is
also seen in the $T$-matrix analysis of the data
\cite{16,bugg}.

\newpage
\begin{center}
Table  1\\
Coupling constants of the glueball decaying to two pseudoscalar mesons,
in the leading\\ and next-to leading terms of $1/N$ expansion.
$\Theta$ is the mixing angle for  $\eta -\eta'$ mesons:
$\eta=n\bar n \cos\Theta-s\bar s \sin\Theta$ and
$\eta'=n\bar n \sin\Theta+s\bar s \cos\Theta$.
\vskip 0.5cm

\begin{tabular}{|c|c|c|c|}
\hline
~       &     ~                &  ~                    &~            \\
~       & Glueball decay       & Glueball decay        &Iden-       \\
~       & couplings in the     & couplings in the      &tity      \\
Channel & leading term of    &next-to-leading term &factor \\
~       &$1/N$ expansion.      &of $1/N$ expansion.    &       \\ ~
& ~                &  ~ &~            \\ \hline ~ & ~ & ~ & ~ \\
$\pi^0\pi^0$ & $G^L$ & 0 & 1/2 \\ ~ & ~ & ~ & ~ \\ $\pi^+\pi^-$ & $G^L$
& 0 & 1  \\ ~ & ~ & ~ & ~ \\ $K^+K^-$ & $\sqrt \lambda G^L$ & 0 & 1 \\
~ & ~ & ~ & ~ \\
$K^0K^0$ & $\sqrt\lambda G^L $ & 0 & 1 \\
~ & ~ & ~ & ~ \\
$\eta\eta$ &
$G^L\left (\cos^2\Theta+ \lambda\sin^2\Theta\right )$
&$2G^{NL}(\cos\Theta-\sqrt{\frac{\lambda}{2}}\sin\Theta )^2$ &
1/2 \\
~ & ~ & ~ & ~ \\
$\eta\eta'$ &
$G^L (1-\lambda)\sin\Theta\;\cos\Theta$
&$2G^{NL}(\cos\Theta-\sqrt{\frac{\lambda}{2}}\sin\Theta)\times$ &1 \\
~&~&$(\sin\Theta+\sqrt{\frac{\lambda}{2}}\cos\Theta)$ & ~\\
~ & ~ & ~ & ~ \\
$\eta'\eta'$ &
$G^L\left(\sin^2\Theta+\lambda\;\cos^2\Theta\right)$
&$2G^{NL}\left(\sin\Theta+\sqrt{\frac{\lambda}{2}}\cos\Theta \right)^2$
	& 1/2 \\ ~ & ~ & ~ & ~ \\
\hline
\end{tabular}
\end{center}

\newpage
\begin{center}
Table 2\\
 Coupling constants of $q\bar q$-meson decaying to
two pseudoscalar mesons in the leading and next-to-leading
terms of the $1/N$ expansion.
$\phi$ is the mixing angle for $n\bar n$ and $s\bar s$ states,
see (1.9).
\vskip 0.5cm
\begin{tabular}{|c|c|c|}
\hline
~      &     ~                        &  ~                             \\
~      & Decay couplings of           &Decay couplings of             \\
~      &  $q\bar q$ mesons            & $q\bar q$ mesons   \\
Channel& in leading term            &in next-to-leading term\\ ~
       & of $1/N$ expansion.          &of $1/N$ expansion.   \\
\hline
~ &~ & ~ \\ $\pi^0\pi^0$ & $g^L\;\cos\phi/\sqrt{2}$& 0  \\ ~ & ~ & ~  \\
$\pi^+\pi^-$ & $g^L\;\cos\phi/\sqrt{2}$ & 0 \\
~ & ~ & ~ \\
$K^+K^-$ & $g^L (\sqrt 2\sin\phi+\sqrt \lambda\cos\phi)/\sqrt 8 $ & 0 \\
~ & ~ & ~ \\
$K^0K^0$ & $g^L (\sqrt 2\sin\phi+\sqrt \lambda\cos\phi)/\sqrt 8 $ & 0 \\
~ & ~ & ~ \\
$\eta\eta$ &
$g^L\left (\cos^2\Theta\;\cos\phi/\sqrt 2+\right .$\hfill
&$\sqrt 2
g^{NL}(\cos\Theta-\sqrt{\frac{\lambda}{2}}\sin\Theta )\times$\hfill\\
~ &\hfill$\left . \sqrt{\lambda}\;\sin\phi\;\sin^2\Theta\right )$ &
\hfill$(\cos\phi\cos\Theta-\sin\phi\sin\Theta)$ \\
~ & ~ & ~  \\
$\eta\eta'$ &
$g^L\sin\Theta\;\cos\Theta\left(\cos\Phi/\sqrt 2-\right .$\hfill
&$\sqrt{\frac 12}
g^{NL}\left [(\cos\Theta-\sqrt{\frac{\lambda}{2}}\sin\Theta)\times
\right .$\\
~& \hfill $\left .\sqrt{\lambda}\;\sin\phi\right ) $ &
\hfill$(\cos\phi\sin\Theta+\sin\phi\cos\Theta)$\\
~&~&$+(\sin\Theta+\sqrt{\frac{\lambda}{2}}\cos\Theta)\times$\hfill\\
~&~&\hfill$\left .(\cos\phi\sin\Theta-\sin\phi\cos\Theta)\right ]$\\
~ & ~ & ~  \\
$\eta'\eta'$ &
$g^L\left(\sin^2\Theta\;\cos\phi/\sqrt 2+\right .$\hfill
&$\sqrt 2
g^{NL} (\sin\Theta+\sqrt{\frac{\lambda}{2}}\cos\Theta)\times$\hfill\\
~&\hfill $\left .\sqrt{\lambda}\;\sin\Phi\;\cos^2\Theta\right)$ &
\hfill$(\cos\phi\cos\Theta+\sin\phi\sin\Theta )$ \\
~ & ~ & ~  \\
\hline
\end{tabular}
\end{center}

\newpage
\centerline{\epsfxsize=11cm \epsfbox{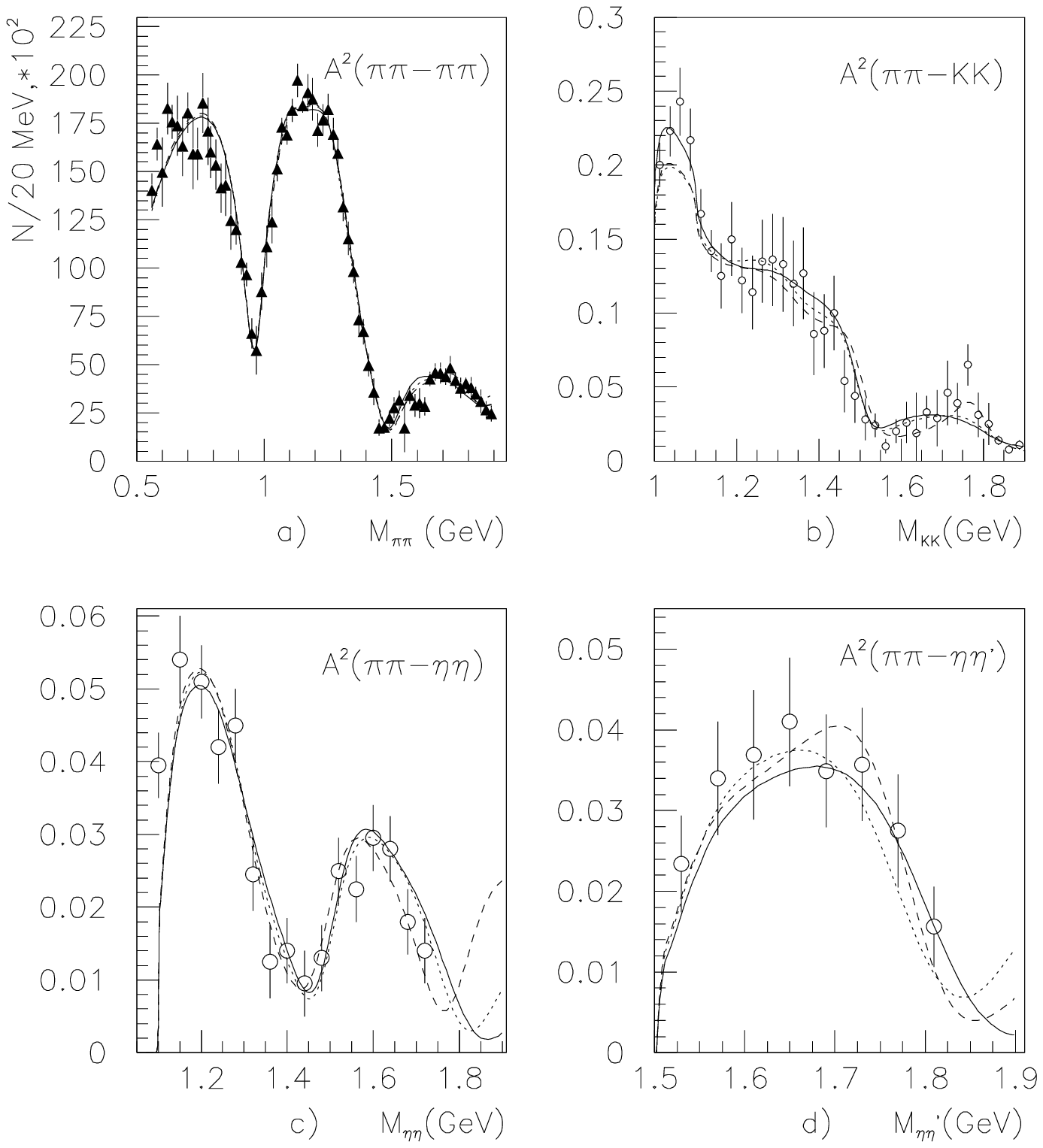}}
Fig. 6.
$S$ wave amplitudes squared and their description in Ref.
 \cite{km1900}: solid curve stands for Solution {\bf II}.
\centerline{\epsfxsize=11cm \epsfbox{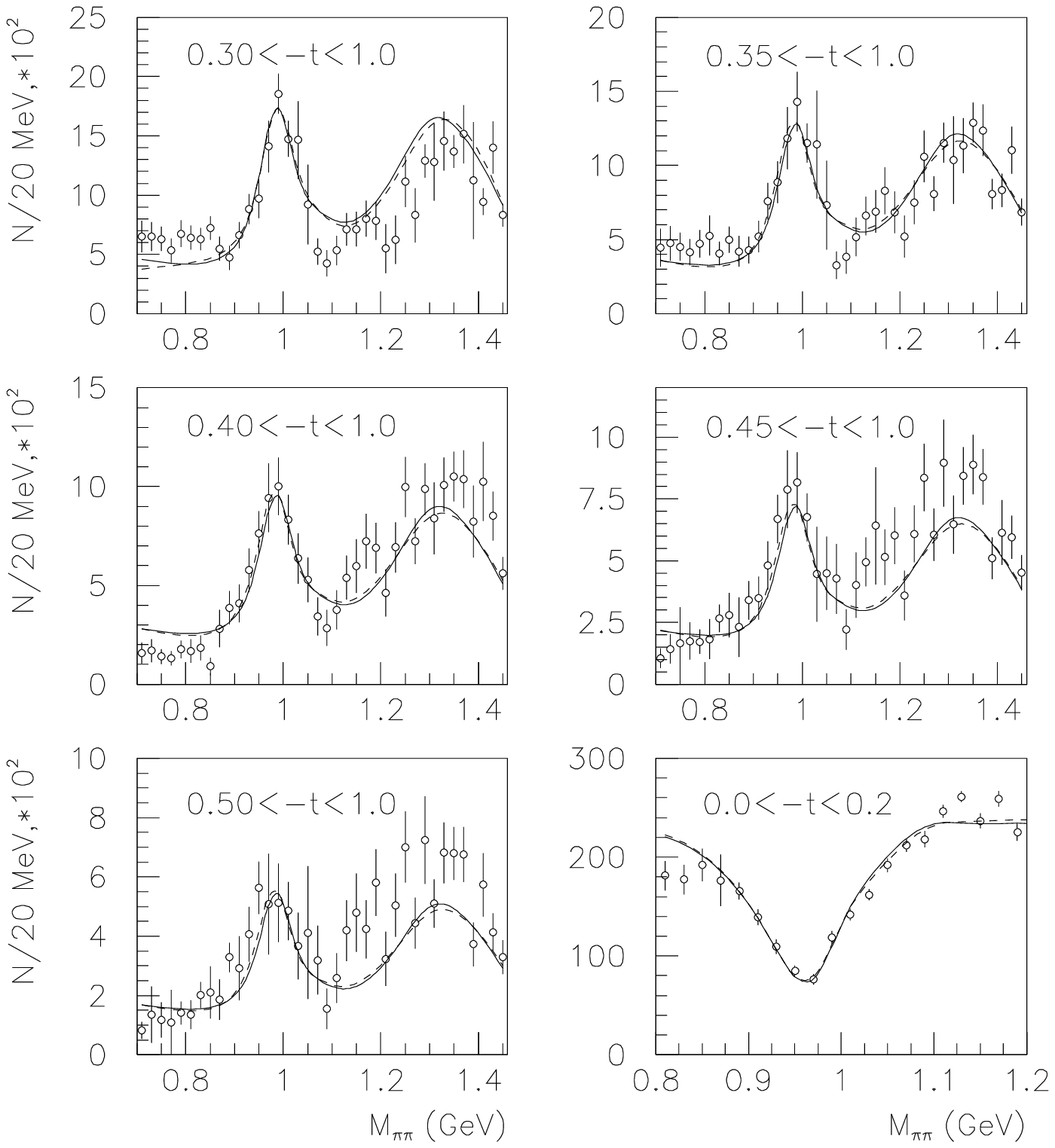}}
Fig. 7.  Event numbers {\it versus}
 invariant mass of the $\pi\pi$-system selected for various
intervals of the momentum transfer squared $t$.
Solid curve stands for Solution {\bf{II}}, dashed one to Solution
 {\bf I}.

\centerline{\epsfxsize=10cm \epsfbox{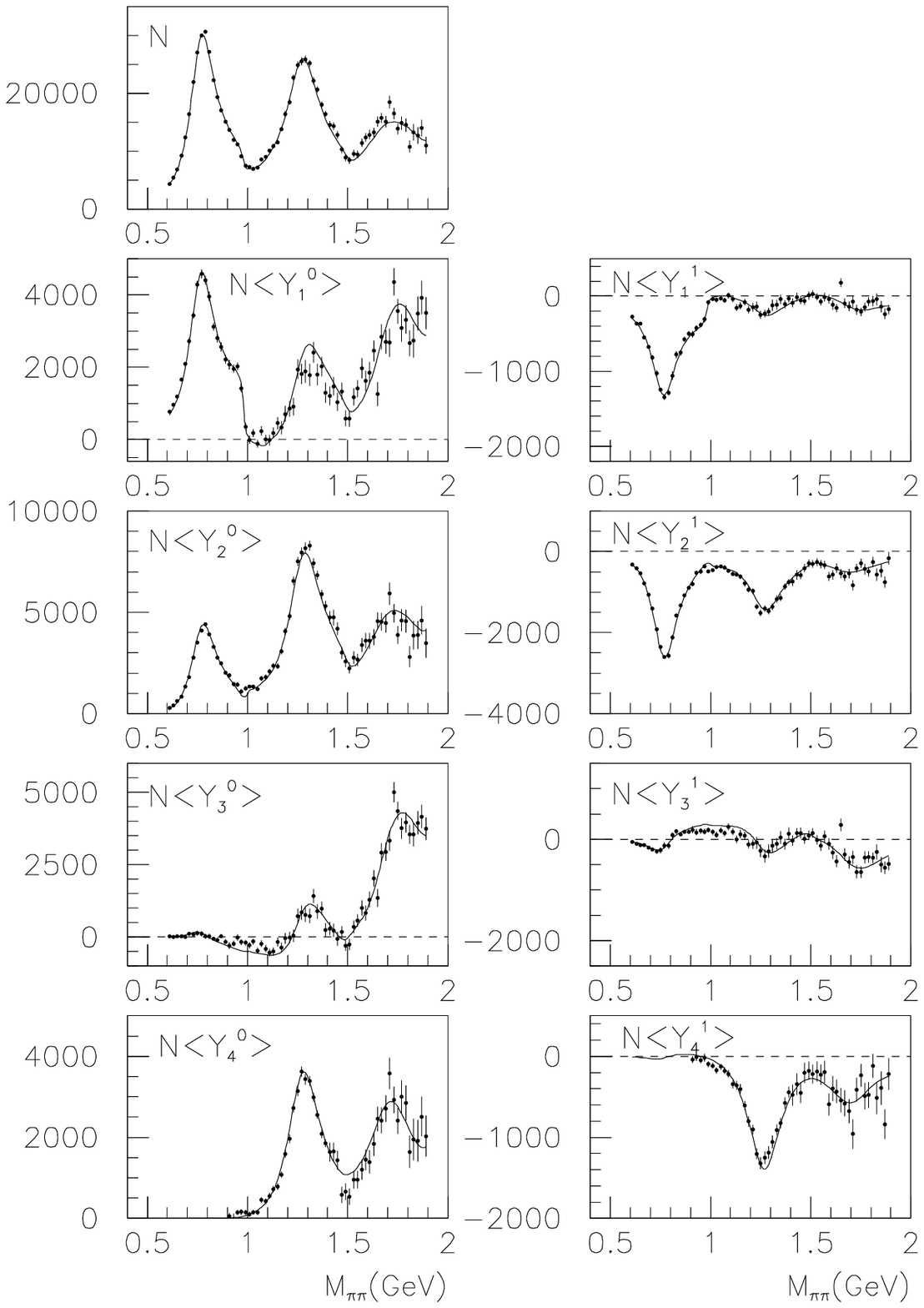}}
Fig. 8. Description of the angle moments
for the  $\pi\pi$ distributions  measured in the reaction
$\pi^- p\to n \pi^-\pi^+$  \cite{cern};
Solution  {\bf II} \cite{km1900}. \\

\centerline{\epsfxsize=11cm \epsfbox{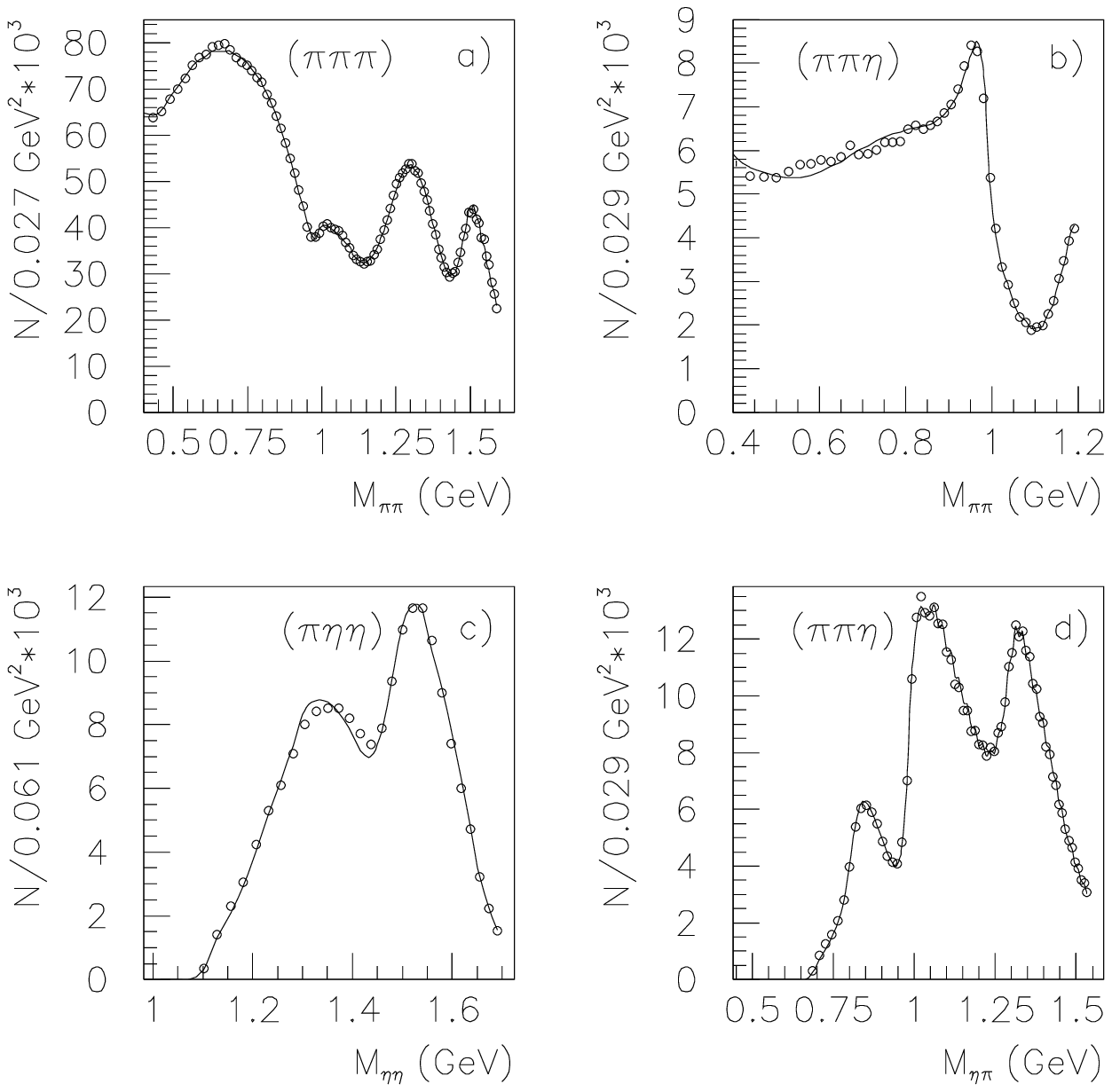}}
Fig. 9. The
 $\pi^0\pi^0$ spectra in the reactions
$p\bar p\to \pi^0\pi^0\pi^0$,
$p\bar p\to \eta\pi^0\pi^0$ and    $\eta\eta$ spectrum in the reaction
 $p\bar p\to \pi^0\eta\eta$ and   $\eta\pi^0$ spectrum
in the reaction $p\bar p\to \pi^0\pi^0\eta$.
Curves correspond to Solution {\bf{II}}  \cite{km1900}.

\centerline{\epsfxsize=11cm \epsfbox{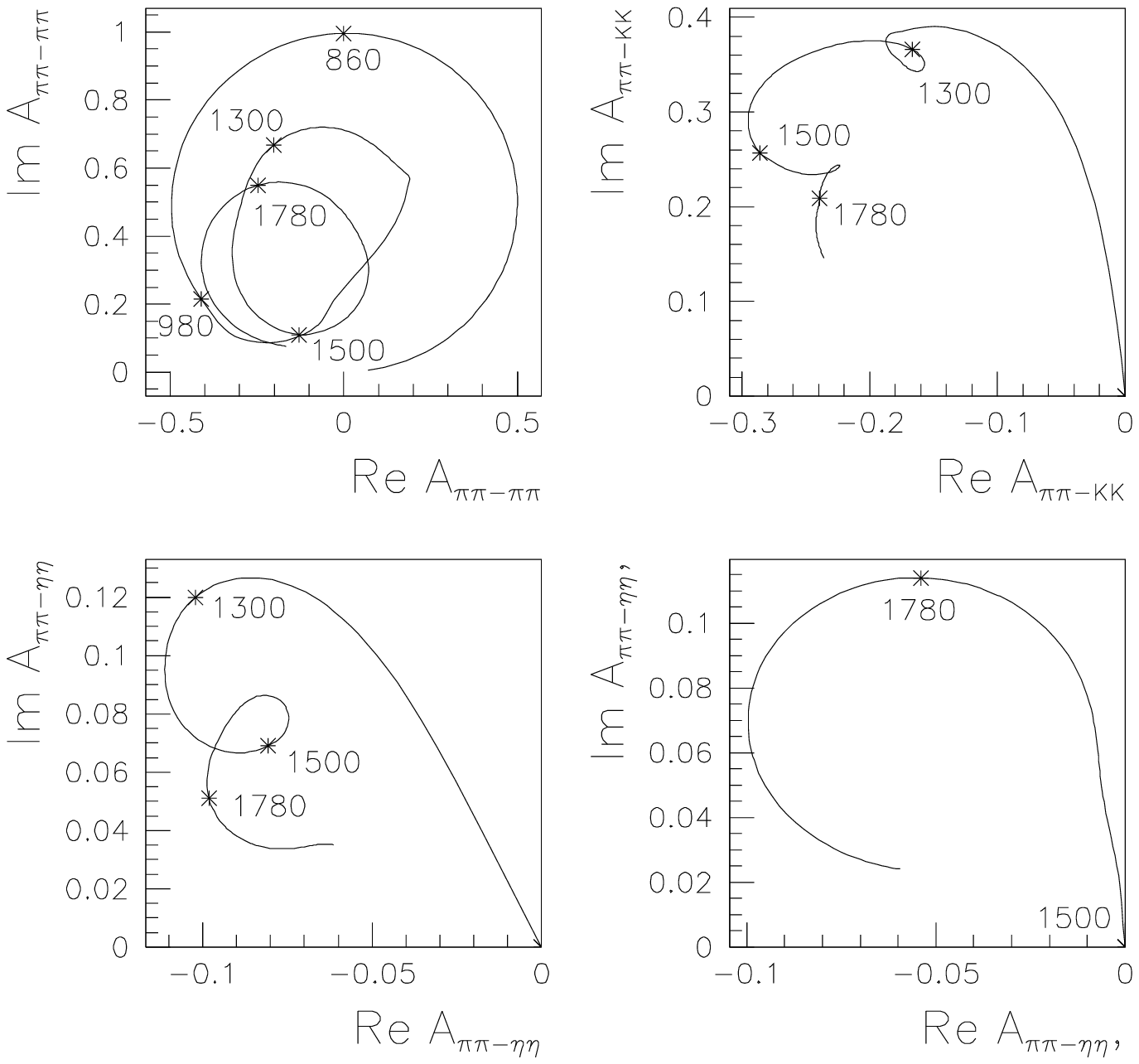}}
Fig. 10.
Argand diagram for the unitary S-wave amplitudes in the reactions
 $\pi\pi\to \pi\pi$,
 $\pi\pi\to K\bar K$,
 $\pi\pi\to \eta\eta$ and
 $\pi\pi\to \eta\eta'$  \cite{km1900}.

\centerline{\epsfxsize=11cm \epsfbox{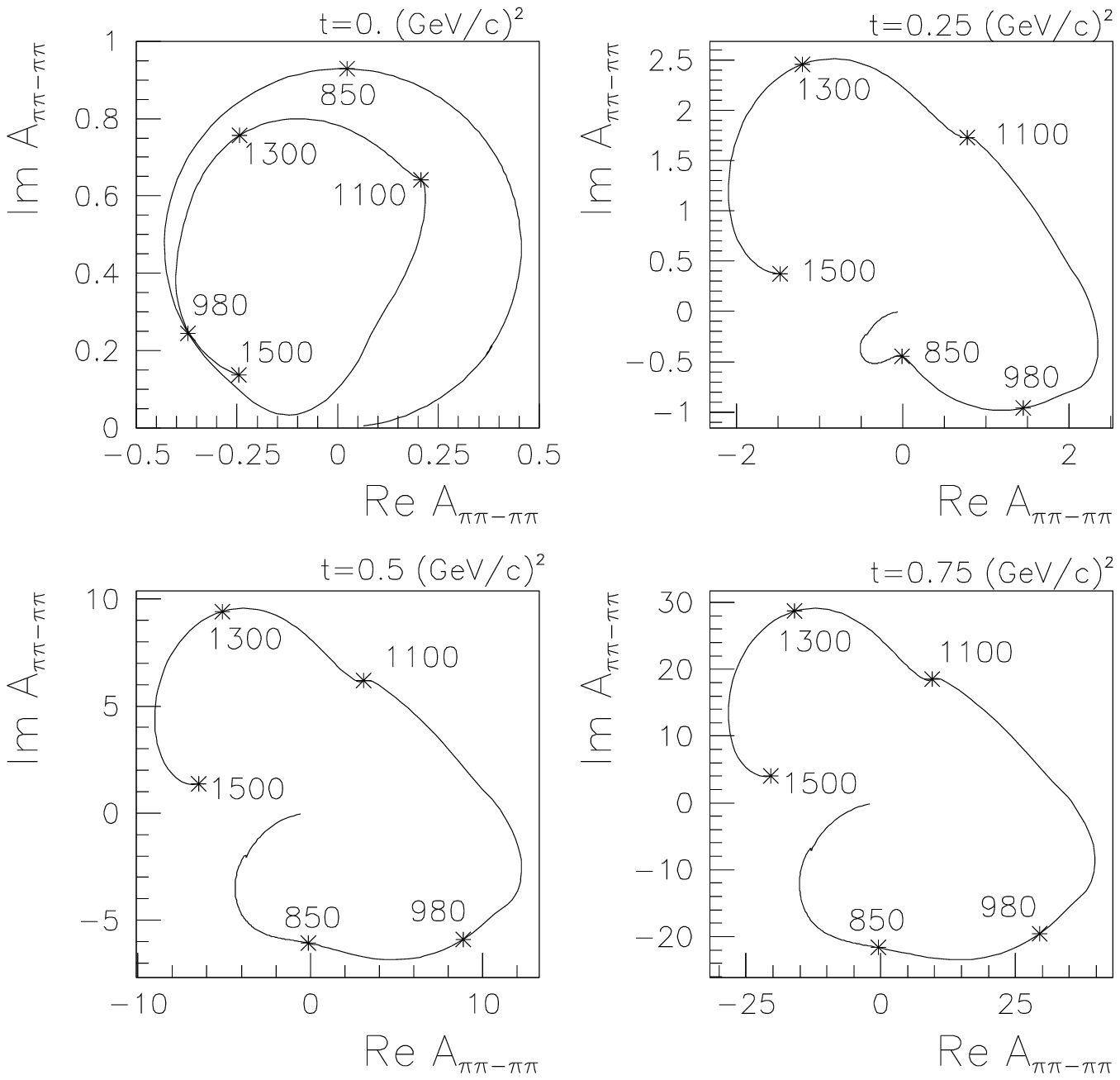}}
Fig. 11.
Argand diagram for the isoscalar S-wave  $\pi\pi(t)\to \pi\pi$
scattering amplitude  \cite{km1900} at different momentum
transfers squared, $t$.

\subsection{$K$-matrix analysis for the $K\pi$ $S$-wave}

Partial--wave analysis of the  $K^- \pi^+$ system for the reaction
$K^- p \to  K^- \pi^+ n$ at  11 GeV/c was carried out in Ref.
  \cite{kpi}, where two alternative solutions (A and B),
 which differ only in the region above 1800 MeV,
were found  for the $S$-wave. In that paper
the $T$-matrix fit for the $K\pi$ $S$-wave was
 performed as well, though independently for the regions
 $850-1600$ MeV and $1800-2100$ MeV.   In the first region, the
resonance $K^*_0(1430)$ was found:
$$ M_R=1429 \pm 9\;
 \mbox{MeV}, \;\;\; \Gamma=  287 \pm 31 \; \mbox{MeV}\;.
\eqno{(3.15)} $$
In the second mass region, Solutions A and B provided
the following parameters for the description of
the resonance $K^*_0(1950)$:
$$Solution\; A
 \qquad\qquad M_R=1934 \pm 28\; \mbox{MeV}, \;\;\; \Gamma= 174 \pm
 98 \; \mbox{MeV}, $$
$$Solution\;B \qquad\qquad M_R=1955 \pm
 18\; \mbox{MeV}, \;\;\; \Gamma= 228 \pm 56 \; \mbox{MeV}.
\eqno{(3.16)} $$
The necessity to improve this analysis is obvious.
First, the mass region $1600-1800$ MeV, where the amplitude varies
quickly, must be included into consideration. As was stressed
 above, it is  well-known  that, due to the strong interference,
the resonance reveals itself not only as a bump
in the spectrum but also as
 a dip or a shoulder;  likewise, the resonances appear in the
$00^{++}$ wave. Second, the interference effects are the source of
ambiguities.  It is worth noting that  ambiguities in $00^{++}$
wave were successfully eliminated
in the analyses of Refs.  \cite{km1900,fullkm},
that was due to the simultaneous fit of different
meson spectra only.  For the wave $\frac{1}{2}0^+$,
the available  data are not copious, hence one may
suspect that the solution found
in Ref.  \cite{kpi} is not unique.

The $K$-matrix re-analysis of the $K\pi\;$ $S$-wave has been
 carried out in Ref. \cite{kmkpi} with a purpose:

$(i)$ to restore the masses and couling constants of the bare states
for the wave $\frac{1}{2} 0^+$, in order to establish the
 $q\bar q$-classification;

$(ii)$ to find out all possible $K$-matrix solutions for the  $K\pi$
 $S$-wave in the mass region up to 2000 MeV.

 The $S$-wave $K \pi$ scattering amplitude extracted from the reaction
$K^- p$ $\to K^-\pi^+ n$ at small momentum transfers is a sum of
two components, with isotopic spins $\frac{1}{2}$ and
 $\frac{3}{2}$:
$$
  A_S= A^{1/2}_S + \frac12
 A^{3/2}_S =\mid A_S \mid e^{i\phi_S}\; ,
\eqno{(3.17)}
$$
where $\mid A_S \mid$ and $\phi_S$  are measurable quantities
 entering the
$S$-wave amplitude \cite{kpi}.  The part of  $S$-wave
amplitude with the isotopic spin $I=3/2$ has a non-resonance
behaviour at the energies under consideration, so it can be parametrised
as follows:
 $$ A_S^{3/2}(s)= \frac{\rho_{K\pi}(s) a_{3/2}(s)}{1- i
\rho_{K\pi}(s) a_{3/2}(s)}\; , \eqno{(3.18)} $$
where $ a_{3/2}(s)$ is a smooth function and  $\rho_{K\pi}(s)$ is
 the $K\pi$ phase space factor.

For the description of the $A^{1/2}_S$ amplitude,  in \cite{kmkpi}
the $3\times  3$ $K$-matrix was used,  with the
following channel notations:
$ 1=K \pi, \;2=K\eta',\; 3=K \pi \pi \pi+ multimeson\; states$.  The
account of the channel $K \eta$ does not influence the data
description, since the transition $K\pi \to K \eta$ is suppressed
\cite{kpi}, that also agrees with quark combinatoric results, see
Table 3.  In Ref. \cite{kmkpi} the fitting to the wave $\frac12 0^+$
was performed with the parametrization of $K_{ab}$ given in (3.1).
The analysed data  for the reaction  $K^- p\to K^- \pi^+ n$ were
extracted with small momentum transfers ($|t|<0.2$ GeV$^2$), and at the
first stage the data were fitted to the unitary amplitude (3.1). At the
next stage, the  $t$-dependence was introduced into the $ K$-matrix
 amplitude. The amplitude $K \pi (t)\to K\pi$ ($\pi (t)$ stands for
virtual pion) is equal to $A^{1/2}_S= K_{1a}(t)\; (I-i\rho
 K)^{-1}_{a1}$:  the parametrization of the matrix $K_{1a}(t)$ is
written in  (3.3).

Coupling constants are determined by the rules of quark combinatorics,
they are presented in Table 3.  In Ref. \cite{kmkpi}
only the leading terms in the $1/N$ expansion were taken into
consideration: in this case,  coupling constants were fixed
by the fit of the $00^{++}$ and $10^{++}$ waves, because $g^L$ is a
common parameter for all the nonet members.

The description of the $\frac120^+$-wave has been performed under
two assumptions, namely, with  two- and three-pole
structure of the wave in the mass region below 2000 MeV.

In Ref. \cite{kpi} two solutions, A and B, were found for the
wave $\frac120^+$; they differ at $M_{\pi K}>1800$ MeV only.
Correspondingly, in Ref. \cite{kmkpi} the two two-pole $K$-matrix
solutions, (A-1) and (B-1), have been obtained. The positions of
amplitude poles are practically the same
for both solutions; they are shown in Eq. (1.16).
The description of data is shown in Fig. 12. The mass of the
first resonance, see (1.16),  does not differ strongly  from that
obtained in \cite{kpi}, see  (3.15), but the width of the resonance
found in the  $K$-matrix solution is twice as less.
This follows from the doubling of poles in the $ K$-matrix solution
due to the correct account of the  $K\eta'$ threshold.
  The mass of the second resonance decreased for the
$K$-matrix solution, as compared to the result of Ref.  \cite{kpi}, in
more than 100 MeV.

The masses of bare kaon states related to the two-pole solution are
given in Eq. (1.17). The mass of lightest state is
$1200^{+60}_{-110}$ MeV, i.e. this scalar kaon is located in the same
mass region as the other scalars, candidates for basic $1^3P_0$
nonet members.

The description of data in the three-pole $K$-matrix fit is shown in
Fig. 13. The region of high masses, $M_{K\pi}>1700$ MeV
 is described by two poles in  Solutions (A-2) and (B-2).
However, the two-pole
structure of the amplitude at $M_{K\pi}>1700$ MeV
did not influence the characteristics of two low--located
 resonances --- they are identical in Solutions  (A-2) and (B-2).
In Solution (B-3) the region $M_{K\pi}<1600$ MeV is described by
two poles. Positions of bare states in Solution  (B-3)
are given in (1.18), the corresponding positions of poles  are given in
Eq. (1.19).

Solid curves in Figs. 12 and 13 represent the description of the
$K \pi$ wave by the unitary amplitude, and dashed lines represent the
fits, where the $t$-dependence of the  $K\pi$ amplitude is taken into
account. It is seen that the  $t$-dependence allows us to get better
description of phase shifts around 1700 MeV.  It should be noted that
in this region as well as in the mass region above 2000 MeV for
 Solution A, the data under consideration violate the unitary limit. It
 is hardly possible that rather strong violation of unitarity
is a consequence of the amplitude $t$-dependence; it is more likely
related to the underestimation of systematic errors in the partial-wave
analysis of Ref. \cite{kpi} in those regions. The inclusion of the
$t$-dependence into fitting procedure does not affect  strongly the
masses of bare states and positions of pole of the amplitude. As a
rule, the masses of bare states found in the $t$-dependent fit are less
in 20-30 MeV than masses obtained in the $t$-independent fits.

\newpage
\begin{center}
Table 3\\
Coupling constants for the transitions $K_0^0\to$ {\it two mesons}
 and $a^-_0\to $ {\it two mesons} in the  leading and next-to-leading
terms of the  $1/N$ expansion.
\vskip 0.5cm
\begin{tabular}{|l|c|c|}
\hline
~ & ~ & ~\\
Channel & Couplings for     & Couplings for \\
        &  leading terms    & next-to-leading terms\\
~ & ~ & ~\\
\hline
~ & ~ &~ \\
$K^+ \pi^-$ & $g^L/2$ & 0 \\
~ & ~ & ~\\
$K^0 \pi^0$ & $-g^L/\sqrt{8}$ & 0\\
~ & ~ & ~\\
$K^0 \eta$  & $(\cos \Theta/\sqrt 2 - \sqrt{\lambda}
 \sin \Theta)g^L/2$ &
$\left (\sqrt 2\cos\Theta -\sqrt\lambda\sin\Theta\right)g^{NL}/2$\\
~ & ~& ~ \\
$K^0 \eta'$  & $(\sin\Theta/\sqrt 2 + \sqrt{\lambda}
 \cos \Theta)\;g^L/2$ &
$\left (\sqrt 2\sin\Theta -\sqrt\lambda\cos\Theta\right)g^{NL}/2$\\
~ & ~ & ~\\
\hline
~ & ~ &~ \\
$K^- K^0$ & $g^L\sqrt{\lambda} /2$ & 0 \\
~ & ~ & ~\\
$\pi^- \eta$  & $g^L\cos \Theta/\sqrt 2 $ &
$\left (\sqrt 2\cos\Theta -\sqrt\lambda\sin\Theta\right)g^{NL}/2$\\
~ & ~& ~ \\
$\pi^- \eta'$  & $g^L\sin\Theta/\sqrt 2 $ &
$\left (\sqrt 2\sin\Theta -\sqrt\lambda\cos\Theta\right)g^{NL}/2$\\
~ & ~ & ~\\
\hline

\end{tabular}
\end{center}

\centerline{\epsfxsize=11cm \epsfbox{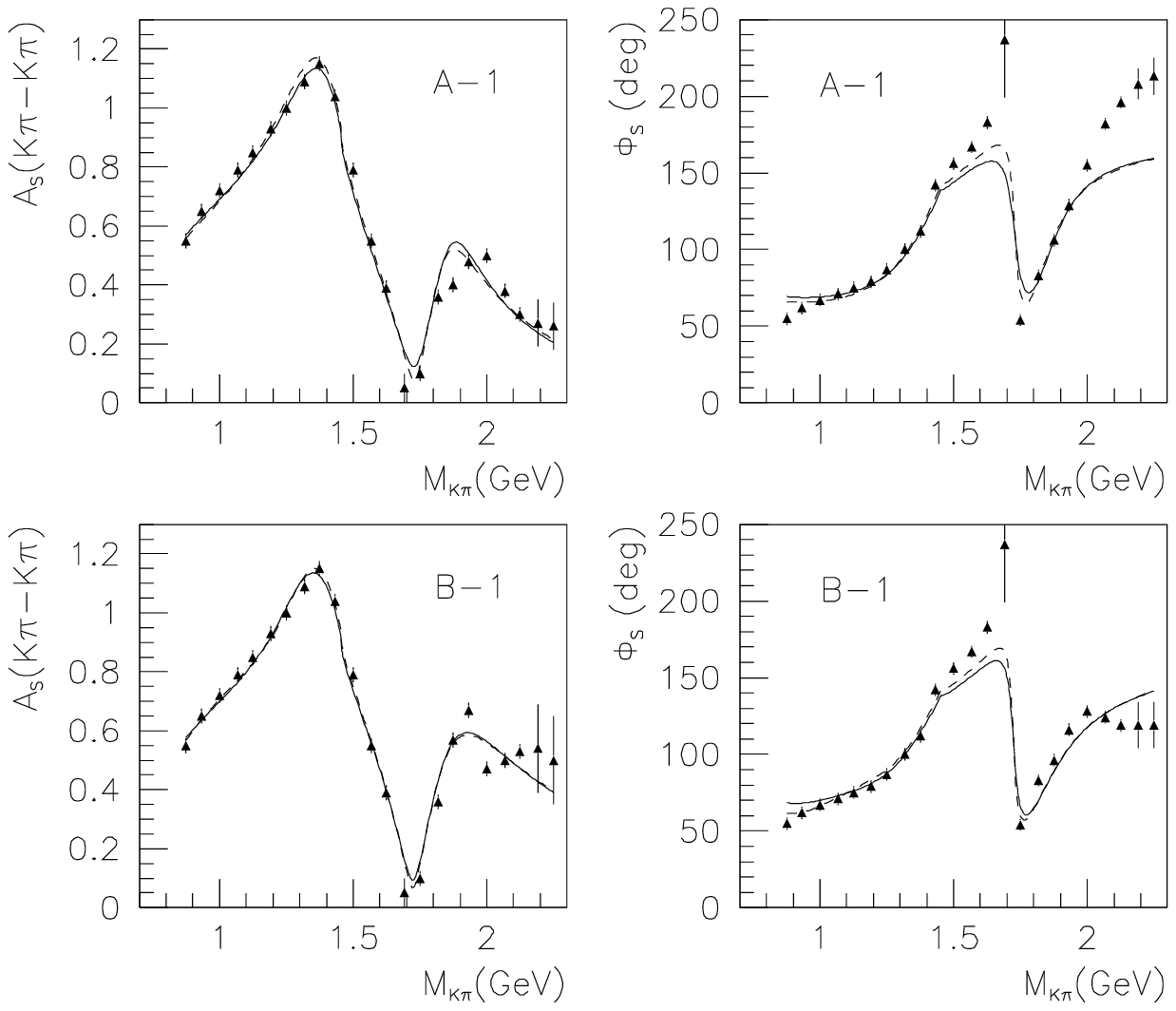}}
Fig. 12. Description of data in Ref. \cite{kpi} in the two-pole
$K$-matrix fit: Solutions (A-1) and (B-1). Solid curves correspond to
the solution found for the unitary amplitude, dashed line stands for
the fit with the $t$-dependent $K$-matrix.

\centerline{\epsfxsize=11cm \epsfbox{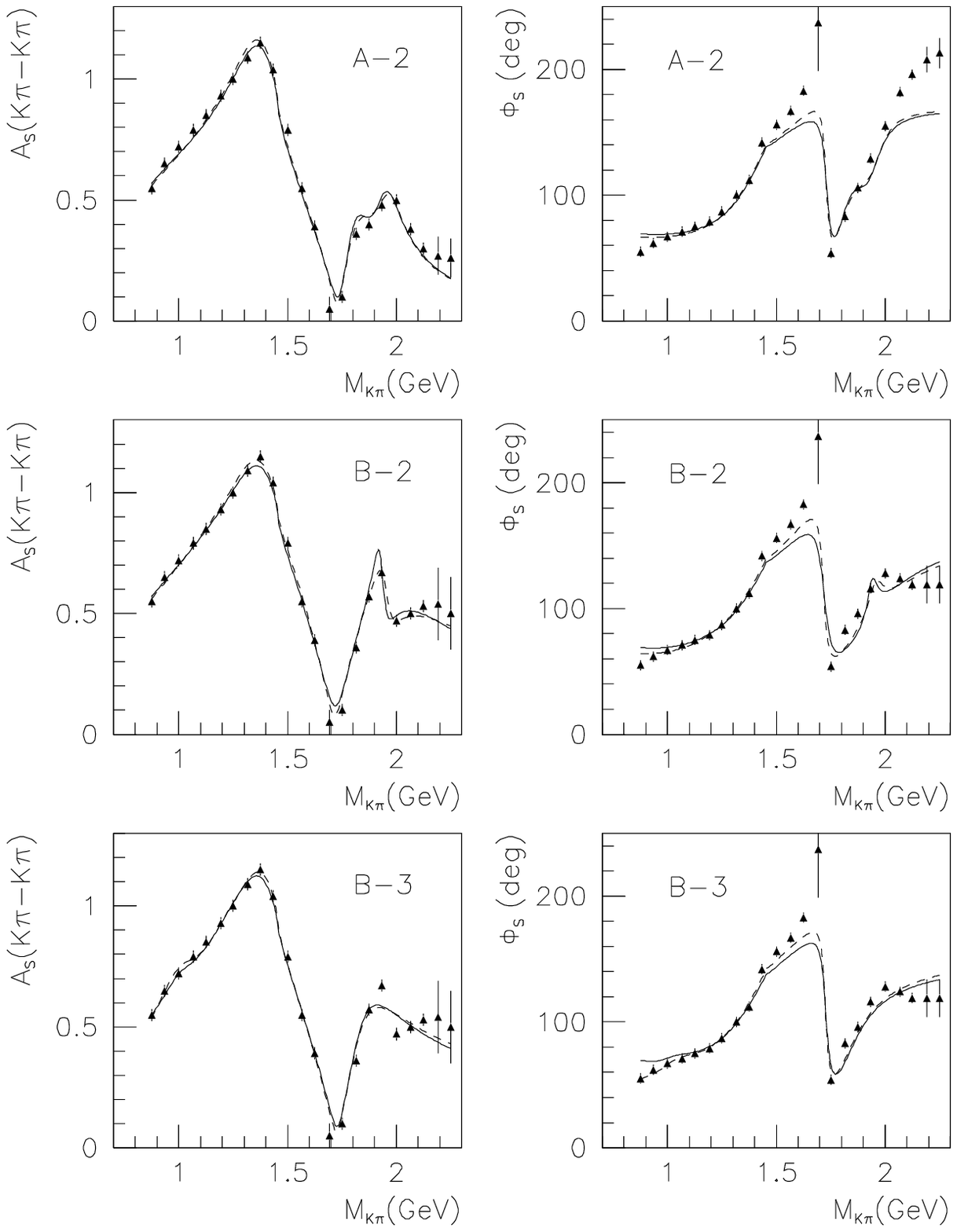}}
Fig. 13. Description of data in Ref. \cite{kpi} in the three-pole
$K$-matrix fit: Solutions (A-2), (B-2) and (B-3).

\section{Propagator matrix: analysis of the $(IJ^{PC}=00^{++})$-wave}

Here we are summing up the results of the analysis of the $00^{++}$
$S$-wave performed in \cite{5,aas}, in terms of propagator matrix
($D$-matrix). The $D$-matrix technique is based on the
dispersion  relation
$N/D$-method, it allows us to reconstruct the amplitude,
which is analytical on the whole complex $s$-plane.
We discuss the effects
 which are due to the resonance overlapping and mixing: mass
shifts and accumulation of widths by one of neighbouring resonances.
The expansion of the physically observed states in a series with
respect to initial (non-mixed) ones is performed.

The investigation is done for the $00^{++}$ wave, though the method
can be easily generalised for other waves using the
technique developped in Refs. \cite{deut,as}.

\subsection{The mixing of two unstable states}

In the case of two resonances, the propagator of the state 1
is determined by the diagrams of Fig.  14a. With
all these processes taken into account, the propagator of the state  1
is equal to:  $$
D_{11}(s)=\left(m_1^2-s-B_{11}(s)-\frac{B_{12}(s)B_{21}(s)}
{m_2^2-s-B_{22}(s)}\right)^{-1}\,.  \eqno{(4.1)} $$
Here $m_1$ and $m_2$  are masses of the input states 1 and 2, and
the loop diagrams $B_{ij}(s)$ are defined by Eq. (2.21), with the
replacement $g^2(s) \rightarrow g_i(s)g_j(s)$.
It is helpful to introduce the propagator matrix  $D_{ij}$,
where the nondiagonal elements $D_{12}=D_{21}$ correspond to the
transitions $1\to 2$ and $2\to 1$ (see Fig.  14b). The matrix reads:
 $$ \hat D=\left|
\begin{array}{ll} D_{11}& D_{12}\\
D_{21}& D_{22}\end{array}\right|
=\frac{1}{(M_1^2-s)(M_2^2-s)-B_{12}B_{21}} \left|\begin{array}{cc}
M_2^2-s,& B_{12}\\
B_{21},& M_1^2-s
\end{array}\right|\,.
\eqno{(4.2)}
$$
Here the following notation is used:
$$
M_i^2=m_i^2-B_{ii}(s)\qquad\qquad i=1,2\;.
\eqno{(4.3)}
$$
Zeros of the denominator of the propagator matrix  (4.2)
define the complex  resonance masses  after the mixing:
$$ \Pi(s)=(M_1^2-s)(M_2^2-s)-B_{12}B_{21}=0\;.  \eqno{(4.4)} $$
Let us denote the complex masses of mixed states by $M_A$ and $M_B$.

Consider a simple model, where the
$s$-dependence of the function $B_{ij}(s)$ near
the points $s\sim M_A^2$ and $s\sim M_B^2$
is assumed to be negligible.
Let $M_i^2$ and $B_{12}$ be constants, then one has:
  $$ M_{A,B}^2=\frac 12 (M_1^2+M_2^2)\pm\sqrt{\frac
14 (M_1^2-M_2^2)^2+ B_{12}B_{21}}\quad.  \eqno{(4.5)} $$
In the case, when the widths of initial resonances 1 and 2 are small
(hence the imaginary part of the transition diagram $B_{12}$ is also
small), the equation (4.5) turns into the standard formula of
quantum mechanics for the split of mixing levels, which
become repulsive as a result of the mixing. Then
 $$ \hat D=\left|\begin{array}{cc}
\frac{\cos^2\theta}{M_A^2-s}+\frac{\sin^2\theta}{M_B^2-s}&
\frac{-\cos\theta\sin\theta}{M_A^2-s}+\frac{\sin\theta\cos\theta}
{M_B^2-s}\\
\frac{-\cos\theta\sin\theta}{M_A^2-s}+\frac{\sin\theta\cos\theta}
{M_B^2-s}&
\frac{\sin^2\theta}{M_A^2-s}+\frac{\cos^2\theta}{M_B^2-s}
\end{array}\right|,
\cos^2\theta=\frac 12+\frac 12\frac{\frac 12(M_1^2-M_2^2)}
{\sqrt{\frac 14(M_1^2-M_2^2)^2+B_{12}B_{21}}}.
$$
$$\eqno{(4.6)}
$$
The states $|A>$ and $|B>$ are  superpositions of initial levels,
 $|1>$ and $|2>$, as follows:
$$
|A>=\cos\theta|1>-\sin\theta|2>\; , \qquad
|B>=\sin\theta|1>+\cos\theta|2>\; .  \eqno{(4.7)} $$

In general,
representation of states $|A>$ and $|B>$ as  superpositions
of initial states is valid, when one cannot neglect
the $s$-dependence of functions  $B_{ij}(s)$ and  their imaginary
parts are not small. Consider the propagator matrix near $s=M_A^2$:
$$
\hat D=\frac{1}{\Pi(s)}\left|\begin{array}{cc} M_2^2(s)-s&B_{12}(s)\\
B_{21}(s)&M_1^2(s)-s
\end{array}\right|
\simeq\,\frac{-1}{\Pi'(M_A^2)(M_A^2-s)} \left|\begin{array}{cc}
M_2^2(M_A^2)-M_A^2& B_{12}(M_A^2)\\
B_{21}(M_A^2)&
M_1^2(M_A^2)-M_A^2\end{array}\right|\; .
\eqno{(4.8)}
$$
In the left-hand side of Eq. (4.8), the singular (pole) terms are
the only  surviving. The matrix determinant in the right-hand side of
(4.8) is equal to zero:
$$
[M_2^2(M_A^2)-M_A^2][M_1^2(M_A^2)-M_A^2]-B_{12}(M_A^2)B_{21}(M_A^2)=0\quad,
\eqno{(4.9)}
$$
This equality follows from Eq. (4.4), which fixes $\Pi(M_A^2)=0$.
It allows us to introduce the complex mixing
angle:
$$
|A>=\cos\theta_A|1>-\sin\theta_A|2>\,.  \eqno{(4.10)} $$
The right-hand
side of Eq. (4.8) can be rewritten with the use of the mixing angle
$\theta_A$, as follows:
$$ \left[\hat D\right]_{s\sim
M_A^2}=\frac{N_A}{M_A^2-s}\left|\begin{array}{cc}
\cos^2\theta_A&-\cos\theta_A\sin\theta_A\\
-\sin\theta_A\cos\theta_A&\sin^2\theta_A\end{array}\right|\,,
\eqno{(4.11)}
$$
where
$$
N_A=\frac{1}{\Pi'(M_A^2)}[2M_A^2-M_1^2-M_2^2], \,
\cos^2\theta_A=\frac{M_A^2-M_2^2}{2M_A^2-M_1^2-M_2^2}, \,
\sin^2\theta_A=\frac{M_A^2-M_1^2}{2M_A^2-M_1^2-M_2^2}\, .
\eqno{(4.12)}
$$
We
remind that in the formula (4.12) the functions $M_1^2(s)$, $M_2^2(s)$
and $B_{12}(s)$ are fixed in the point $s=M_A^2$. In the case under
consideration, when the angle $\theta_A$ is a complex
quantity, the values $\cos^2\theta_A$  and
$\sin^2\theta_A$ do not determine the probability of states
$|1>$ and $|2>$ in $|A>$; indeed, the values $\sqrt{N_A}\cos\theta_A$
and $-\sqrt{N_A}\sin\theta_A$ are the transition amplitudes
$|A>\rightarrow |1>$ and $|A>\rightarrow |2>$.  Therefore, the
corresponding probabilities are equal to $|\cos\theta_A|^2$ and
$|\sin\theta_A|^2$.

In order to analyse the content of the state $|B>$, an analogous
expansion of the propagator matrix should be done near the point
 $s=M_B^2$.  After introducing
$$
|B>=\sin\theta_B|1>+\cos\theta_B|2>\,, \eqno{(4.13)} $$
we have the following expression for  $\hat D$
in the vicinity of the second pole $s=M_B^2$:
$$
\left[\hat D\right]_{s\sim M_B^2}=\frac{N_B}{M_B^2-s}\left|
\begin{array}{ll}\sin^2\theta_B &\cos\theta_B\sin\theta_B\\
\sin\theta_B\cos\theta_B &\cos^2\theta_B\end{array}\right|\,,
\eqno{(4.14)} $$
where
$$
N_B=\frac{1}{\Pi'(M_B^2)}\left[2M_B^2-M_1^2-M_2^2\right], \;
\cos^2\theta_B=\frac{M_B^2-M_1^2}{2M_B^2-M_1^2-M_2^2},\;
\sin^2\theta_B=\frac{M_B^2-M_2^2}{2M_B^2-M_1^2-M_2^2}.
\eqno{(4.15)}
$$
In Eq. (4.15) the functions $M_1^2(s)$, $M_2^2(s)$ and $B_{12}(s)$
are fixed in the point $s=M_B^2$.

If $B_{12}$ depends weakly on $s$ and one can neglect this dependence,
the angles  $\theta_A$ and $\theta_B$ coincide. But in general
they are different. So the formulae for the propagator matrix differ
from the standard approach of quantum mechanics by this very point.

Another distinction is related to the type of the level shift
afforded by mixing, namely, in quantum mechanics the levels "repulse"
each other from
 the mean value $1/2(E_1+E_2)$ (see also Eq.  (4.5)). Generally,
the equation (4.4) can cause both the "repulsion" of masses squared
from the mean value, $1/2(M_1^2+M_2^2)$, and the "attraction".

The scattering amplitude in the one-channel case is defined by the
following expression:
$$ A(s)=g_i(s)D_{ij}(s)g_j(s)\,.  \eqno{(4.16)}
$$
In the multichannel case, $B_{ij}(s)$ is a sum of loop diagrams:
$$ B_{ij}(s)=\sum_{n}B_{ij}^{(n)}(s)\,, \eqno{(4.17)} $$
$B_{ij}^{(n)}$ is a loop diagram in the channel
 $n$ with vertex functions $g_i^{(n)}$, $g_j^{(n)}$
and phase space
factor $\rho_n$.  Partial scattering amplitude in the channel
 $n$ is equal to:
$$
A_n(s)=g_i^{(n)}(s)D_{ij}(s)g_j^{(n)}(s)\,.
\eqno{(4.18)}
$$

\subsection{The overlapping of a large number of resonances:
construction of propagator matrix}

Consider the construction of the  propagator matrix
$\hat D$ for arbitrary number of resonances. The matrix elements,
 $D_{ij}$, describe the transition from the input state $i$
(with the bare propagator $(m_i^2-s)^{-1}$) to the state $j$. They obey
the system of linear equations as follows:
$$ D_{ij}=D_{ik}B_{kj}(s)(m^2_j-s)^{-1}+\delta_{ij}(m_j^2-s)^{-1}\;,
\eqno{(4.19)}
$$
where  $B_{ij}(s)$ is the loop diagram for the transition $i \to j$
 and  $\delta_{ij}$ is the Kronecker symbol.
 Let us introduce the diagonal propagator matrix $\hat d$ for input
states :
$$ \hat d=diag\left ( (m_1^2-s)^{-1} ,(m_2^2-s)^{-1}
,(m_3^2-s)^{-1} \cdots \right )\,.  \eqno{(4.20)} $$
Then the system of linear equations  (4.19) can be
rewritten in the matrix form as follows:
$$ \hat D= \hat
D \hat B \hat d +\hat d\;.  \eqno{(4.21)} $$
One obtains:
$$ \hat
D=\frac{I}{(\hat d^{-1}-\hat B)}\;.  \eqno{(4.22)} $$
The matrix  $\hat d^{-1}$ is diagonal, hence
$\hat D^{-1}=(\hat d^{-1}-\hat B)$ is of the form:
$$ \hat
D^{-1}=\left |\begin{array}{cccc} M_1^2-s & -B_{12}(s) & -B_{13}(s)
&\cdots\\ -
B_{21}(s) &M_2^2-s &  -B_{23}(s) &\cdots\\ -B_{31}(s)
&-B_{32}(s) &M_3^2-s &  \cdots\\
\vdots & \vdots & \vdots & \vdots
\end{array}\right |\;, \eqno{(4.23)} $$
where $M^2_i$ is defined by Eq. (4.3).  Inversing this matrix,
we obtain a full set of elements $D_{ij}(s)$:
$$
D_{ij}(s)=\frac{(-1)^{i+j}\Pi_{ji}^{(N-1)}(s)}{\Pi^{(N)}(s)}\;.
\eqno{(4.24)}
$$
Here $\Pi^{(N)}(s)$ is the determinant of the matrix $\hat D^{-1}$,
and $\Pi_{ji}^{(N-1)}(s)$ is a matrix supplement to the element
$[\hat D^{-1}]_{ji}$, i.e. the  matrix
$\hat D^{-1}$ with the excluded $j$-th line and $i$-th column.

The zeros of $\Pi^{(N)}(s)$ define the poles of the
propagator matrix which correspond to physical resonances formed by
the mixing. We denote the complex resonance masses as:
$$
s=M_A^2\,,\quad M_B^2\,,\quad M_C^2\,, \ldots \eqno{(4.25)} $$
Near the point $s=M_A^2$, one can leave in
the propagator matrix the leading pole term only.
This means that  the free term in Eq. (4.21)
can be neglected, so we get a system of homogeneous equations:
$$ D_{ik}(s)\left ( \hat
d^{-1}-\hat B \right )_{kj}=0\,.  \eqno{(4.26)} $$
The solution of this system is defined  up to the
normalization factor, and it does not depend on the initial index
$i$. Then the elements of the propagator matrix may be written in a
factorized form as follows:
$$ \left[\hat D^{(N)}\right]_{s\sim M_A^2}=\frac{N_A}{M_A^2-s}\cdot
\left|\begin{array}{llll}\alpha_1^2,&\alpha_1\alpha_2,&
\alpha_1\alpha_3, & \ldots\\
\alpha_2\alpha_1,&\alpha_2^2,&\alpha_2\alpha_3,& \ldots\\
\alpha_3\alpha_1,&\alpha_3\alpha_2,&\alpha_3^2,& \ldots\\
\ldots &  \ldots & \ldots & \ldots\end{array}\right|\,,
\eqno{(4.27)}
$$
where $N_A$ is the normalization factor chosen to satisfy
the  condition:
$$
\alpha_1^2+\alpha_2^2+\alpha_3^2+\ldots+\alpha_N^2=1\,.
\eqno{(4.28)}
$$
The constants $\alpha_i$ are the normalized amplitudes for the
transitions:
{\it resonance~A} $\rightarrow$ {\it state} $i$. The probability
to find the state $i$ in the physical resonance $A$ is equal to:
$$
w_i=|\alpha_i|^2\;.
\eqno{(4.29)}
$$
Analogous representation
of the propagator matrix can be also done in the
vicinity of other poles:
$$ D_{ij}^{(N)}(s\sim
M_B^2)=N_B\frac{\beta_i\beta_j}{M_B^2-s}\,,\qquad D_{ij}^{(N)}(s\sim
M_C^2)=N_C\frac{\gamma_i\gamma_j}{M_C^2-s}\, \qquad \cdots.
\eqno{(4.30)}
$$
Coupling constants satisfy  normalization conditions similar
to that of Eq. (4.28):
$$ \beta_1^2+\beta_2^2+\ldots+\beta_N^2=1\,,\qquad
\gamma_1^2+\gamma_2^2+\ldots+\gamma_N^2=1\,,\qquad\cdots\,.
\eqno{(4.31)}
$$
In general case, however, there is no completness condition for the
inverse expansion:
$$ \alpha_i^2+\beta_i^2+\gamma_i^2+\ldots\neq 1\,.
\eqno{(4.32)}
$$
For two resonances, it means  that  $\cos^2\Theta_A+
\sin^2\Theta_B\neq 1$. Still, let us remind that the equality in the
inverse expansion, which is  relevant to the completness condition,
appears in the models, where the $s$-dependence of loop diagrams
is neglected, see Eqs. (4.5)--(4.7).

\subsection{Full resonance overlapping: the accumulation of widths of
neighbouring resonances by one of them  }

Let us consider two examples which  describe the idealized situation of
a full overlapping of two or three resonances. In these examples, the
effect of accumulation of widths of neighbouring resonances by one of
them can be seen in its original untouched form.

a) \underline{Full overlapping of two resonances.}

For the simplicity sake, let $B_{ij}$ be a weak $s$-dependent
function, and Eq. (4.5) may be used. We define:
$$
M_1^2=M_R^2-iM_R\Gamma_1\,,\qquad M_2^2=M_R^2-iM_R\Gamma_2\,,
\eqno{(4.33)}
$$
and
$$
ReB_{12}(M_R^2)=P\int_{(\mu_1+\mu_2)^2}^{\infty}\frac{ds'}{\pi}
\frac{g_1(s')g_2(s')\rho(s')}{s'-M_R^2} \to 0\,.
\eqno{(4.34)}
$$
It is possible that $ReB_{12}(M_R^2)$ can be equal to zero at positive
$g_1$ and $g_2$, if the contribution from the integration
region $s'<M_R^2$ cancels the contribution from the region
$s'>M_R^2$.  In this case
$$ B_{12}(M_R^2) \to
ig_1(M_R^2)g_2(M_R^2)\rho(M_R^2)= iM_R\sqrt{\Gamma_1\Gamma_2}\,.
\eqno{(4.35)}
$$
Substituting Eqs. (4.33)--(4.35) into Eq. (4.5), one has:
$$
M_A^2  \to M_R^2-iM_R(\Gamma_1+\Gamma_2)\, \qquad
M_B^2  \to M_R^2\,.
\eqno{(4.36)}
$$
Therefore, after mixing, one of the states accumulates the widths of
primary resonances, $\Gamma_A \to \Gamma_1+\Gamma_2$, and another state
becomes  quasi-stable particle, with $\Gamma_B \to 0$.

b) \underline{Full overlapping of three resonances.}

Consider the equation
$$  \Pi^{(3)}(s)=0
\eqno{(4.37)}
$$
at the same approximation as in the above example.
Correspondingly, we put:
$$  Re B_{ab}(M_R^2) \to 0\, , (a\neq b); \qquad
M_i^2=M_R^2-s-iM_R\Gamma_i=x-i\gamma_i\;.
\eqno{(4.38)}
$$
A new variable, $x=M_R^2-s$, is used, and we denote $M_R\Gamma_i=
\gamma_i$. Taking account of  $B_{ij}B_{ji}=-\gamma_i\gamma_j$ and
$B_{12}B_{23}B_{31}=-i\gamma_1\gamma_2\gamma_3$,
we can re-write the equation (4.37) as follows:
$$
x^3+x^2(i\gamma_1+i\gamma_2+i\gamma_3)=0\,.
\eqno{(4.39)}
$$
Therefore, at full resonance overlapping, one obtains:
$$
M_A^2 \to M_R^2-iM_R(\Gamma_1+\Gamma_2+\Gamma_3)\; , \qquad
M_B^2\to M_R^2\; , \qquad  M_C^2\to M_R^2\,.
\eqno{(4.40)}
$$
The resonance  $A$ has accumulated the widths of three primary
resonances, and the states $B$ and $C$  became
quasi-stable and degenerate.

\subsection{The resonances $f_0(1300)$, $f_0(1500)$,
$f_0(1530^{+90}_{-250})$ and $f_0(1780)$}

The $K$-matrix analysis provides a basis for the
investigation of mixing phenomenon in the scalar sector.
Propagator matrix technique used at the next stage  of the
analysis allows us to restore correctly the contribution from the real
parts of loop diagrams, $B_{ij}(s)$, thus having correctly calculated
contributions of input states to the formation of physical resonances.

The resonance mixing in the region
1200-1600 MeV may be considered in the  two-channel approximation, for
the quark-hadron duality justifies the analysis with the use of
  quark channels $n\bar n$ and $s\bar s$. Correspondingly,
$$
B_{ij}(s)=\cos\varphi_i\cos\varphi_jB_{ij}^{(n\bar n)}(s)+\sin\varphi_i
\sin\varphi_jB_{ij}^{(s\bar s)}(s)\,,
\eqno{(4.41)}
$$
where $i$, $j$ run over $1,2,3,4$, with the following notations
for  states: $1=1^3P_0(n\bar n \; rich)$, $2=2^3P_0(n\bar n
\; rich)$, $3=gluonium$ and $4=2^3P_0(s\bar s \; rich)$.
The quark states are usually described by the light cone
variables. Then,
$$ B_{ij}^{(n\bar
n)}(s)=\frac{1}{(2\pi)^3}\int_{0}^{1}\frac{dx}{x} \int
d^2k_\bot\frac{g_i(s')g_j(s')}{s'-s}2(s'-4m^2)\, .  \eqno{(4.42)} $$
Here $s'=(m^2+k_\bot^2)/x(1-x)$ and $m$ is the mass of nonstrange
quark. The factor  $2(s'-4m^2)$ appears due to the presence of the
quark spin variables:  $Tr[(\hat k+m)(-\hat p+\hat
k+m)]=2(s'-4m^2)$.  Analogous expression, with the replacement $m \to
m_s$, determines $B_{ij}^{(s\bar s)}(s)$.

The simplest parametrization of vertex function for the transition
$input \; state \;i \to quarks$ is:
$$ g_i(s)= \gamma_i\; \sqrt[4] s
\;\left [ \frac{k_i^2+\sigma_i}{k^2+\sigma_i} - d_i
\frac{k_i^2+\sigma_i}{k^2+\sigma_i+h} \right ];\\
\eqno{(4.43)}
$$
Here $k^2=s/4 -m^2$ and $k_a^2=m_a^2/4 -m^2$,
where $m$ is the constituent quark mass equal to 350 MeV for
nonstrange quark and 500 MeV for strange one, and $m_a$ is the mass
of input state.

For the first state, $1^3P_0(n\bar n \; rich)$,  and for the gluonium
we put $d_1=d_3=0$. The second state is the radial excitation,
 $2^3P_0 (n\bar n \; rich)$, and its wave function is
orthogonal to the ground state.
This means that real part of the function  $B_{12}(s)$ must
tend to zero at  $s$ close to resonance masses. Such an
orthogonalization has been performed at the point  $\sqrt s=1.5$
GeV, thus determining the value of the coefficient $d_2$.
Vertex functions for  the same nonet  members are equal to each other,
so $g_2(s)=g_4(s)$.

Parameters $m_a$, $\gamma _a$, $h$ and $\sigma_a$ are defined by
the masses
and widths of physical resonances. However, the mass  $m_a$
can be approximately fixed by the  $K$-matrix pole:
 $\mu_a^2 \simeq m_a^2 -Re\; B_{aa}(\mu_a^2)$.  It should be underlined
that $m_3$ is the mass of a pure gluonium which is a subject of the
 Lattice QCD.

Positions of amplitude poles and the masses of input states found in
Ref.  \cite{aas} by fitting the $00^{++}$ wave are shown in Table 4.
Relative weight of the primary state in the physical resonance
$A$ is defined by Eq. (4.29):
calculated in such a way the probabilities
$W_a$ for the resonances under investigation are shown in Table 4.

As was stressed above, in order to make  comparisons with the QCD
calculations, one should separate the contributions from large and
small distances, that is, to take into account the short-range
interaction component, $r<r_0 \sim R_{confinement}$, and eliminate the
contribution from large $r$.  Therefore, in the calculation of masses
which might be compared with the results of the QCD-motivated models,
we should make a replacement in the amplitude of the $00^{++}$  wave
as follows:  $$ B_{ab}(s)\rightarrow Re\bar B_{ab}(s,k^2_0)
=P\int\limits_{4m^2+4k^2_0}^{\infty}\frac{ds'}{\pi}\; \frac{g_a(s')\rho
(s')g_b(s')}{s'-s}\;2(s'-4m^2)\; .  \label{11} \eqno{(4.44)} $$ The
poles of the amplitude re-determined in this way provide the masses
which are related to the interaction at $r<1/k_0$.  One must compare
with the quark model results the masses obtained with a cutting, of the
order of $k_0^2\sim 0.125$ (GeV/c)$^2$, that corersponds to the account
of the quark interaction at $r\le 1\; fm\sim R_{confinement}$.

For Solution I we get (the values are given in GeV):
$$
\begin{tabular}{lcccc}
~ &$1^3P_0(s\bar s\; rich)$     & $1^3P_0(n\bar n\; rich)$   &
  $2^3P_0(n\bar n\; rich)$ & $2^3P_0(s\bar s\; rich)$ \\
~ \\ $m(k_0^2=0)=\mu_a^{bare}$    & 0.720 & 1.360 & 1.577 & 1.791 \\
  $m(k_0^2=0.125)$           & 0.730 & 1.340 & 1.560 & 1.780 \\
  $m(k_0^2\rightarrow\infty)=m_a$ &  --  & 1.457 & 1.536 & 1.750
  \end{tabular}
\eqno{(4.45)}
$$
In Solution II:
$$
\begin{tabular}{lcccc}
~ &$1^3P_0(s\bar s\; rich)$     & $1^3P_0(n\bar n\; rich)$   &
  $2^3P_0(n\bar n\; rich)$ & $2^3P_0(s\bar s\; rich)$ \\
$m(k_0^2=0)=\mu_a^{bare} $ & 0.720 &   1.357 & 1.585 & 1.734 \\
$m(k_0^2=0.125)$   & 0.735 &   1.340 & 1.570 & 1.725 \\
$m(k_0^2\to\infty)=m_a$ &  --   &   1.107 & 1.566 & 1.702
\end{tabular}
\eqno{(4.46)}
$$

The lightest $q\bar q$-state, $f_0^{bare}(720)$, has not been included
into the mixing machinary in Refs. \cite{5,aas}. In Eqs.
(4.45) and (4.46) the mass corrections for this state have been
evaluated as $ m_a(k_0^2)\simeq m_a^2-Re\;\bar B_{aa}(m_a^2,k_0^2)$.
This approximate equality is due to a comparative smallness of      the
nondiagonal loop diagrams.

Equations (4.45) and (4.46) prove that the values
$m_a(k_0^2=0.125$ GeV$^2$) slightly differ from $\mu_a^{bare}$, while
the differences from the input masses $m_a$ can be significant.
This means that the  $K$-matrix analysis provides approximately correct
meson characteristics which may be compared with the  quark
model results. On the contrary, one should compare with the results of
Lattice QCD the values of input masses $m_a$, which can  noticeably
differ both from the masses of bare states, $\mu_a^{bare}$, and from
those of real resonances.

\subsection{Dynamics of glueball mixing with $q\bar q$-states}

To find out the glueball mixing with  $q\bar q$-
states, let us make a replacement in loop diagrams of the propagator
matrix:
$$
g_a(s)\rightarrow\xi g_a(s)\,,
\eqno{(4.47)}
$$
with the factor $\xi$ changing in the interval $0\leq\xi\leq 1$. At
$\xi \to 0$ the mixing is switched off,
 and the amplitude has the poles at $s \simeq m_a^2$.
Figure 15 demonstrates the position of poles at different $\xi$
for Solutions I and II.  With increasing $\xi$, the poles move from the
real axis to the lower part of the complex plane. Let us discuss in
detail the dynamics of the pole movement for Solution II.

At $\xi=0.1-0.5$ the glueball state of Solution II is mainly mixed
with the state $2^3P_0(n\bar n\,rich)$, while at $\xi=0.8-1.0$
the mixing with the state $1^3P_0(n\bar n\,rich$ becomes important.
As a result, the state, which is  the glueball descendant, sunk
rather deeply into complex plane, acquiring the mass $M=1450-i450$ MeV,
and the glueball component of this broad resonance is 47\%.  Likewise,
in Solution  I  the broad resonance is the gluon descendant as well.

 The hypothesis about   strong mixing of the gluonium with the
$q\bar q$-states was raised formerly. But the attempts to restore
quantitatively the picture of mixing within the
standard quantum mechanics  approach  failed, for two
phenomena have been lost:

(1) The  $q\bar q/glueball$ mixing described by the $D$-matrix may lead
not only to the repulsion of levels, that follows from the standard
quantum mechanical approach, but to the attraction of levels as well.
The latest effect  is caused by the presence of imaginary parts of the
loop diagrams $B_{ab}$, and it is important that $Im \,B_{ab}$ is not
small near 1500 MeV.

(2) The resonance overlapping leads to the repulsion of poles located
on the imaginary axis of masses, and one resonance accumulates the
widths of the others.

This very type of the mixing occurred at 1500 MeV, and a large
width of one of the resonances is its inevitable consequence.

It is also natural that the broad resonance itself is
the gluonium  descendant, for the gluonuim mixes
without any significant suppression with the nearby  $q\bar
q$-states, both of them being dominantly nonstrange ones.

\newpage
\begin{center}
Table 4\\
\vskip 0.3cm
Masses and mixing angles of the input states, \\
the content of physical states and positions of poles of the
$00^{++}$-amplitude (masses in GeV).
\vskip 0.3cm
\begin{tabular}{|c|c|c|c|c|}
\hline \multicolumn{5}{|c|}{Solution I}  \\ \hline
~ & $1^3P_0(n\bar n\; rich)$ & $2^3P_0(n\bar n\; rich)$ &
Gluonium & $2^3P_0(s\bar s\; rich)$\\
 ~ &$\phi_1=18^\circ$ &$\phi_2=-6^\circ$  & $\phi_3=25^\circ$ &
$\phi_4=84^\circ$\\
~ & $m_1=1.457$&$m_2=1.536$ &$m_3=1.230$ &$m_4=1.750$ \\
\hline
$W[f_0(1300)]$ & 32\% & 12\%  & 55\% & 1\%\\
$1.300-i0.115$   & ~ &~ & ~& \\
$W[f_0(1500)]$ & 25\% & 70\%  & 3\% & 2\%\\
$1.500-i0.065$   & ~ &~ & ~& \\
$W[f_0(1530)]$ & 44\% & 24\%  & 27\% & 4\%\\
$1.450-i0.450$   & ~ &~ & ~& \\
$W[f_0(1780)]$ & 1\% &  1\%   &  0\%   & 98\%\\
$1.780-i0.085$   & ~ &~ & ~& \\
\hline
\multicolumn{5}{|c|}{Solution II}  \\
\hline
~ & $1^3P_0(n\bar n\; rich)$ & $2^3P_0(n\bar n\; rich)$ &
Gluonium & $2^3P_0(s\bar s\; rich)$\\
~ &$\phi_1=18^\circ$ &$\phi_2=35^\circ$  & $\phi_3=25^\circ$ &
$\phi_4=-55^\circ$\\
~&$m_1= 1.107$&$m_2=1.566$ &$m_3=1.633$ &$m_4= 1.702$ \\
\hline
$W[f_0(1300)]$ & 35\% & 26\%  & 38\% & 0.4\%\\
$1.300-i0.115$   & ~ &~ & ~& \\
$W[f_0(1500)]$ & 1\% & 64\%  & 35\%  & 0.4\%\\
$1.500-i0.065$   & ~ &~ & ~& \\
$W[f_0(1530)]$ & 12\% & 41\%  & 47\% & 0.3\%\\
$1.450-i0.450$   & ~ &~ & ~& \\
$W[f_0(1780)]$ & 0.1\%  &  0.2\%   & 0.2\%   & 99.5\%\\
$1.750-i0.100$   & ~ &~ & ~& \\
\hline
\end{tabular}
\end{center}

\newpage
\centerline{\epsfxsize=11cm \epsfbox{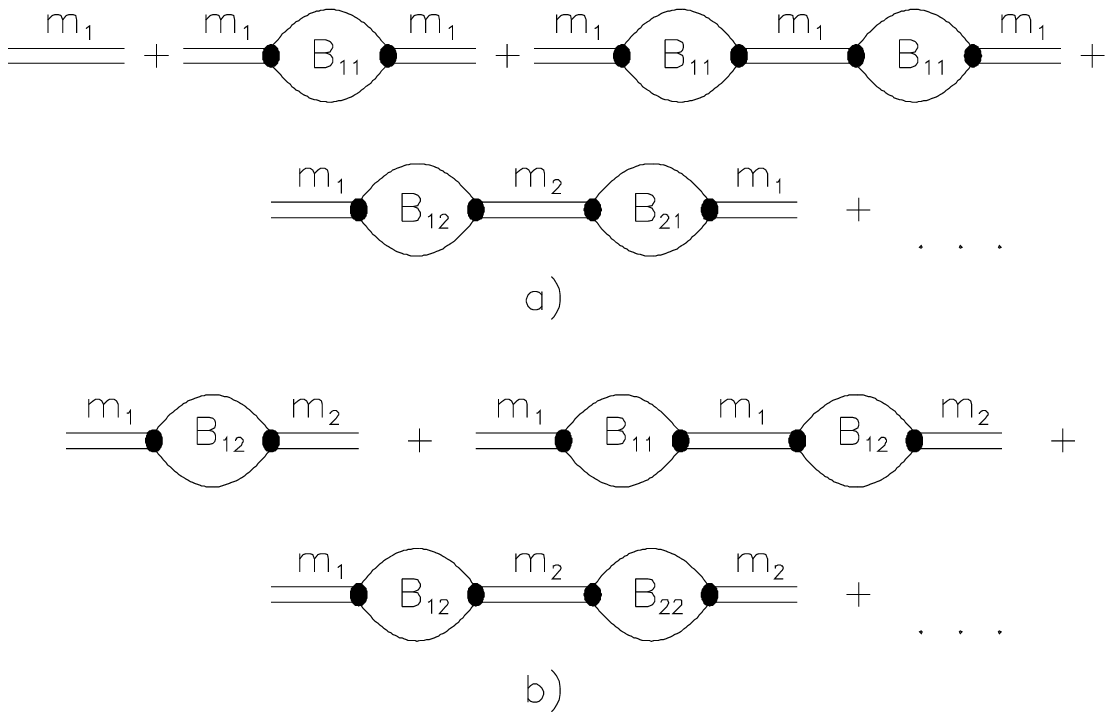}}
Fig. 14. Diagrams describing the propagation
functions $D_{11}$ (a) and $D_{12}$ (b) for
 the interaction of two bare states.

\centerline{\epsfxsize=11cm \epsfbox{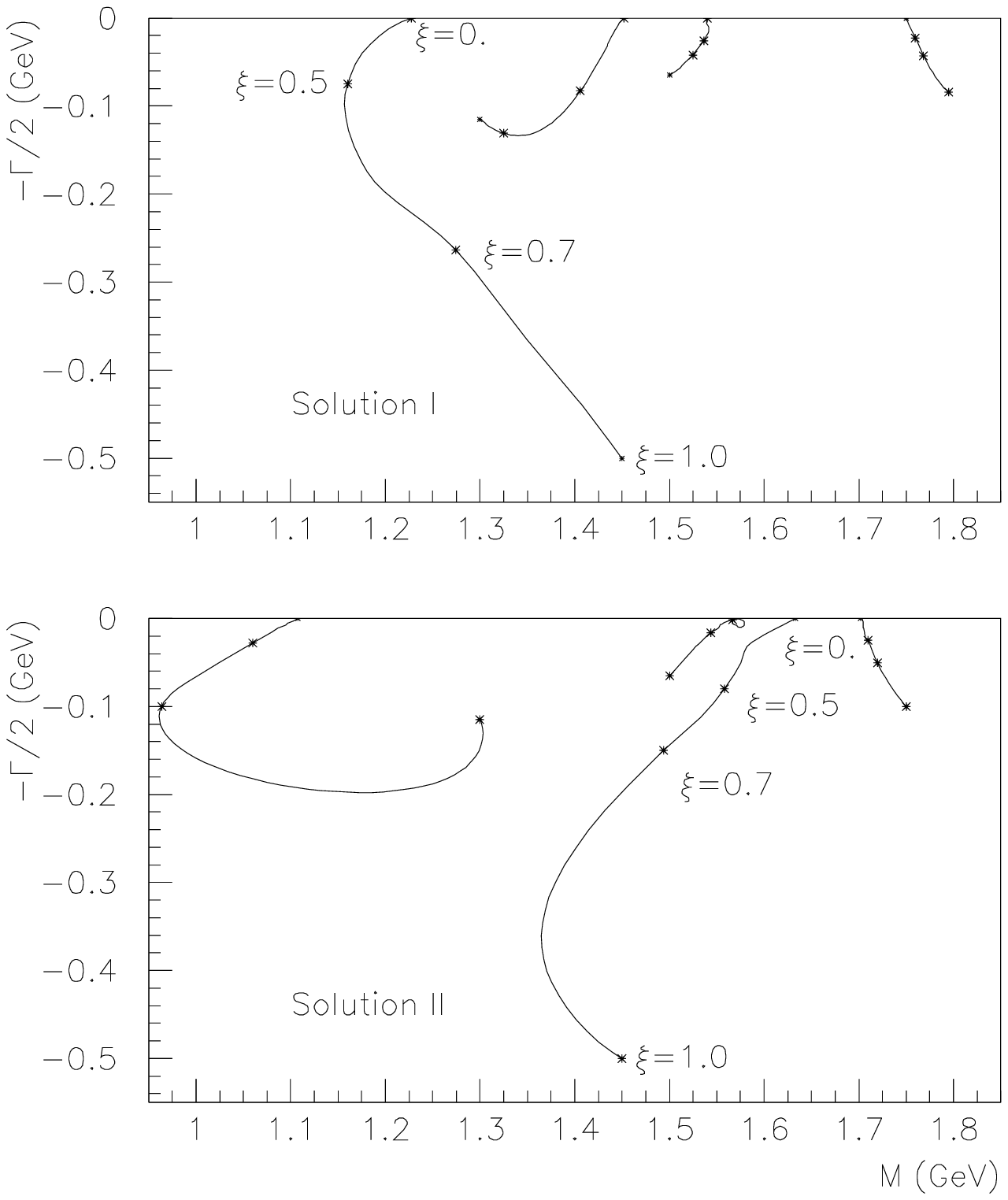}}
Fig. 15. Variation of  the resonance pole positions plotted on the
complex-$M$ plane  at different coupling constants $g_a\to
\xi g_a$.

\section{Conclusion}

Deconfinement of quarks from the excited
$q\bar q$-levels is going in two stages: \\
1) Unavoidable production of $q\bar q$-pairs, which form
two or more white states (hadrons).\\
2) The flying away of the produced hadrons, their interaction and, as a
result, the mixing of neighbouring  $q\bar q$-levels that leads to the
formation of a broad locking state, which plays the role of a
dynamical barrier for  neighbouring levels.

That is  the $K$-matrix analysis together with the dispersion $N/D$
method, which are summoned to decypher the second stage of the
 deconfinement. Analysis of $00^{++}$-wave carried out in the
$K$-matrix or propagator matrix techniques  demonstrated that the
lightest scalar glueball, being near the states  $1^3P_0q\bar q$ and
$2^3P_0q\bar q$, turned after the mixing into a broad state, with
$\Gamma/2 \simeq 500$ MeV. This broad state $f_0(1530^{+90}_{-250})$
carries about a half of the scalar gluonium component, while the other
components of the broad resonance are $1^3P_0q\bar q$ and $2^3P_0q\bar
q$.

It looks like
the waves $00^{-+}$ and  $02^{++}$
behave similarly \cite{dvb}, that allows us to
believe that physics of  highly excited states is tightly
connected with the study of broad states, being an
important and inevitable step in  the search for exotic hadrons and
the investigation of confinement phenomenon.

\section*{} Thanks are due to L.G. Dakhno and V.A. Nikonov for help.
This investigation is supported by RFBR grant N  96-02-17934
and the INTAS-RFBR grant N 95-0267.

\end{document}